\pdfoutput=1
\documentclass[a4 paper, 10pt,twoside]{article}
\usepackage{xcolor}
\usepackage{graphicx}
\usepackage{latexsym}
\usepackage{url}
\usepackage{xcolor}
\usepackage{amsmath, amsthm}

\def\aap{A\&A\,  }

\def\aj{AJ  }
\def\apj{ApJ\,  }
\def\apjl{ApJ Letters,  }

\def\jcap{Journal of Cosmology and Astroparticle Physic  } 
 
\def\jrasc{JRASC  } 
\def\mnras{MNRAS\,  }
\def\pasj{PASJ\,  }
\def\pasp{PASP  }
\def\pasa{PASA  }

\def\prd{Phys. Rev. D   }
\def\zap{Zeitschrift fur Astrophysik} 
 
\begin{document}
\def\2F1{_2F_1}
\def\h0units{\mathrm{km\,s^{-1}\,Mpc^{-1}}}
\def\cunits{\mathrm{km\,s^{-1}}}
\def\lunits{\mathrm{erg\,s^{-1}}}
\newcommand{\om}{\Omega_{\rm M}}
\newcommand{\zerom}{\Omega_{\rm M0}}
\newcommand{\fizero}{\Omega_{\rm \phi\,0}}
\newcommand{\ok}{\Omega_K}
\newcommand{\ola}{\Omega_{\Lambda}}
\newcommand{\ode}{\Omega_{DE}}
\newcommand{\dl}{D_{\rm{L}}}
\newcommand{\dlsette}{D_{\rm{L,7}}}
\newcommand{\dlnum}{D_{\rm{L,num}}}
\newcommand{\mstar}{\ensuremath{m_{B}^\star}\,}
\newcommand{\da}{D_{\rm{A}}}
\def\sun{\hbox{$\odot$}}
\title
{
Sparse formulae  
for the distance modulus in cosmology
}
\vspace{2pc}
\author     {Lorenzo  Zaninetti  \\
Physics Department,
 via P. Giuria 1,\\ I-10125 Turin, Italy\\ 
Email: l.zaninetti@alice.it}
\maketitle
\begin{abstract}
We review  the   distance modulus   in twelve different cosmologies:
the $\Lambda$CDM model,
the wCDM model, 
the Cardassian model,
the flat case,
the $\phi$CDM cosmology,
the Einstein--De Sitter model,
the modified Einstein--De Sitter model,
the simple~GR  model,
the flat expanding model, 
the Milne  model,
the plasma model
and 
the modified tired light model.
The above distance moduli are processed
for three  different compilations of supernovae
and a supernovae + GRBs compilation:
Union 2.1,
JLA,
the Pantheon
and 
Union 2.1 
+ 59 GRBs.
For each of the 48 analysed cases we report 
the relative cosmological parameters, 
the chi-square, the reduced chi-square, the AIC 
and  the $Q$ parameter.
The  angular distance as function of the redshift 
for five  cosmologies
is reported in the framework
of the minimax approximation.
\end{abstract}

\vspace{2pc}
\vspace{2pc}

\section{Introduction}

At the moment of writing, the determination 
of the Hubble constant  is oscillating  
between  a low value 
as derived by the Planck collaboration \cite{Planck2018}, 
$H_0= (67.4 \pm 0.5)\h0units$,
and an high value,
$H_0= (74.03 \pm 1.42)\h0units$,
as  measured  on 
70 long-period Cepheids in the Large Magellanic Cloud (LMC)\cite{Riess2019}.
The  above difference is referred to as the Hubble constant tension
\cite{DiValentino2020} and takes the value
of $4.4 \sigma$.  It fixes an acceptable  interval 
for the evaluation of $H_0$.
The  number of
supernovae (SNs) of type Ia
for which  the distance modulus is available
has grown with time:
34  SNs   in the sample which produced  evidence
for the   accelerating universe  \cite{Riess1998},
580 SNs   in  the Union 2.1 compilation \cite{Suzuki2012},
740 SNs  in
the joint light-curve analysis (JLA)  \cite{Betoule2014},
and  
1048 SNs in the Pantheon sample \cite{Jones2018,Scolnic2018}.
The availability of SN compilations allows  testing
old and new cosmological models.
We select some of them among others:
cosmological relativity in five
spatial dimensions \cite{Oliveira2016},
an improvement of the Einstein--De Sitter cosmology \cite{Gupta2018},
the $f(R)$ gravity with additional logarithmic
corrections \cite{Amarzguioui2006,Odintsov2019},
influence of the detection of gravitational waves
on a
definitive theory of gravity \cite{Corda2009},
the derivation of the value 
of the Hubble constant as $H_0=(70.5\pm 0.5)\h0units$
in the framework of the dark energy cosmology \cite{Lin2019} 
and
the deduction of the parameters for Starobinsky gravity \cite{Camlibel2020}.
This paper reviews, in Section \ref{section_cosmologies},
old and new distance moduli in twelve cosmologies.
Then  
Section \ref{section_results} processes 
the analysed cosmologies in four compilations 
of SNs.

\section{Different cosmologies}

In the following we analyse twelve cosmologies.
A useful introduction  to the distances  
in cosmology can be found in \cite{Hogg1999}.

\label{section_cosmologies}

\subsection{The standard cosmology}

In $\Lambda$CDM cosmology
the {\em Hubble
distance\/} $D_{\rm H}$
is defined as
\begin{equation}
\label{eq:dh}
D_{\rm H}\equiv\frac{c}{H_0}
\quad ,
\end{equation}
where $c$ is the speed of light and $H_0$ is the Hubble constant.
We then introduce a first parameter
 $\om$
\begin{equation}
\om = \frac{8\pi\,G\,\rho_0}{3\,H_0^2}
\quad ,
\end{equation}
where $G$ is the Newtonian gravitational constant and
$\rho_0$ is the mass density at the present time.
A second parameter is $\ola$
\begin{equation}
\ola\equiv\frac{\Lambda\,c^2}{3\,H_0^2}
\quad ,
\end{equation}
where $\Lambda$ is the cosmological constant,
see \cite{Peebles1993}.
Once $\ola$ and $H_0$  are   found 
the numerical value of the cosmological constant 
is derived,   $\Lambda\approx 1.2 \frac{1}{m^2}$.

The two previous parameters are connected with the
curvature $\ok$ by
\begin{equation}
\om+\ola+\ok= 1
\quad .
\end{equation}
The  comoving distance, $D_{\rm C}$,  is
\begin{equation}
D_{\rm C} = D_{\rm H}\,\int_0^z\frac{dz'}{E(z')}
\label{integralez_lcdm}
\quad ,
\end{equation}
where $E(z)$ is the `Hubble function'
\begin{equation}
\label{eq:ez}
E(z) = \sqrt{\om\,(1+z)^3+\ok\,(1+z)^2+\ola}
\quad .
\end{equation}
The above integral cannot be done in  analytical terms, 
except for the case of $\ola=0$, but the Pad\'e approximant, see
Appendix \ref{appendixa}, allows to derive
the approximated indefinite integral, see equation (\ref{integral22}).

The approximate definite integral for (\ref{integralez_lcdm})
is therefore
\begin{equation}
D_{\rm C,2,2} = D_{\rm H}\,
\Big (
F_{2,2}(z;a_0,a_1,a_2,b_0,b_1,b_2)  -
F_{2,2}(0;a_0,a_1,a_2,b_0,b_1,b_2)
\Big )
\label{integralez22}
\quad ,
\end{equation}
where $F_{2,2}$ is equation (\ref{integral22}).
The transverse comoving distance $D_{\rm M}$ is
\begin{equation}
D_{\rm M} = \left\{
\begin{array}{ll}
D_{\rm H}\,\frac{1}{\sqrt{\ok}}\,\sinh\left[\sqrt{\ok}\,D_{\rm C}/D_{\rm H}\right] & {\rm for}~\ok>0 \\
D_{\rm C} & {\rm for}~\ok=0 \\
D_{\rm H}\,\frac{1}{\sqrt{|\ok|}}\,\sin\left[\sqrt{|\ok|}\,D_{\rm C}/D_{\rm H}\right] & {\rm
for}~\ok<0
\end{array}
\right .
\end{equation}
and the approximate transverse comoving distance $D_{\rm
M,2,2}$ computed with the Pad\'e approximant   is
\begin{equation}
D_{\rm M,2,2} = \left\{
\begin{array}{ll}
D_{\rm H}\,\frac{1}{\sqrt{\ok}}\,\sinh\left[\sqrt{\ok}\,
D_{\rm C,2,2}/D_{\rm H}\right] & {\rm for}~\ok>0 \\
D_{\rm C,2,2} & {\rm for}~\ok=0 \\
D_{\rm H}\,\frac{1}{\sqrt{|\ok|}}\,\sin\left[\sqrt{|\ok|}\,D_{\rm C,2,2}/D_{\rm H}\right] & {\rm for}~\ok<0
\end{array}
\right .
\end{equation}

The Pad\'e approximant for the  luminosity distance
is
\begin{equation}
D_{\rm L,2,2} = (1+z)\,D_{\rm M,2,2}
\label{luminositydistancepade}
\quad ,
\end{equation}
and the Pad\'e approximant for the distance modulus, $(m-M)_{2,2}$,
is
\begin{equation}
(m-M)_{2,2} =25 +5 \log_{10}(D_{\rm L,2,2})
\quad .
\label{distmod_lcdm}
\end{equation}
As  a consequence, $M_{2,2}$, the  absolute
magnitude of the Pad\'e approximant, is
\begin{equation}
M_{2,2} = m -25 -5 \log_{10}(D_{\rm L,2,2})
\quad .
\label{absmagz}
\end{equation}

\begin{figure}
\includegraphics[width=6cm]{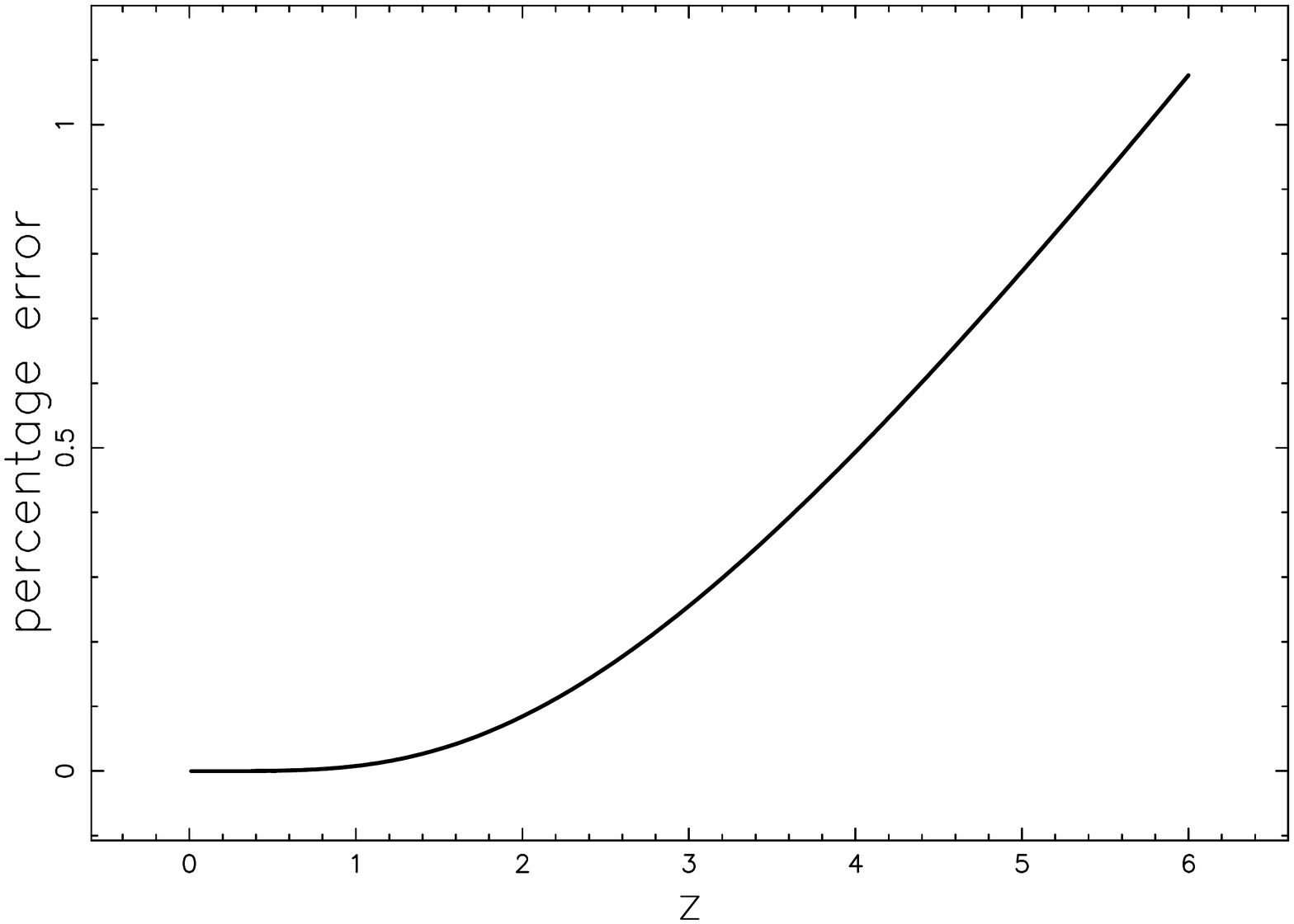}
\caption{
Percentage error  of $(m-M)_{2,2}$ in respect
to the numerical value
with  data as in Table \ref {chi2_union21}. 
}
\label{errore_pade}
\end{figure}

The expanded  version of the Pad\'e
approximant distance modulus
is
\begin{eqnarray}
(m-M)_{2,2} =    \nonumber \\
25+5\,{\frac {1}{\ln  \left( 10 \right) }\ln  \left( {\frac {c \left(
1+z \right) }{H_{{0}}\sqrt {{\it \ok}}}\sinh \left( 1/2\,{\frac {
\sqrt {{\it \ok}}A}{{b_{{2}}}^{2}\sqrt {4\,b_{{0}}b_{{2}}-{b_{{1}}}
^{2}}}} \right) } \right) }
\quad ,
\label{distancemodulusexplicit}
\end{eqnarray}
with
\begin{eqnarray}
A=
\ln  \left( {z}^{2}b_{{2}}+zb_{{1}}+b_{{0}} \right) a_{{1}}b_{{2}}
\sqrt {4\,b_{{0}}b_{{2}}-{b_{{1}}}^{2}}
\nonumber  \\
-\ln  \left( {z}^{2}b_{{2}}+zb_
{{1}}+b_{{0}} \right) a_{{2}}b_{{1}}\sqrt {4\,b_{{0}}b_{{2}}-{b_{{1}}}
^{2}}
\nonumber \\
-\ln  \left( b_{{0}} \right) a_{{1}}b_{{2}}\sqrt {4\,b_{{0}}b_{{2
}}-{b_{{1}}}^{2}}
\nonumber  \\
+\ln  \left( b_{{0}} \right) a_{{2}}b_{{1}}\sqrt {4\,
b_{{0}}b_{{2}}-{b_{{1}}}^{2}}+2\,a_{{2}}zb_{{2}}\sqrt {4\,b_{{0}}b_{{2
}}-{b_{{1}}}^{2}}
\nonumber \\
+4\,\arctan \left( {\frac {2\,zb_{{2}}+b_{{1}}}{
\sqrt {4\,b_{{0}}b_{{2}}-{b_{{1}}}^{2}}}} \right) a_{{0}}{b_{{2}}}^{2}
-2\,\arctan \left( {\frac {2\,zb_{{2}}+b_{{1}}}{\sqrt {4\,b_{{0}}b_{{2
}}-{b_{{1}}}^{2}}}} \right) b_{{1}}a_{{1}}b_{{2}}
\nonumber \\
-4\,\arctan \left( {
\frac {2\,zb_{{2}}+b_{{1}}}{\sqrt {4\,b_{{0}}b_{{2}}-{b_{{1}}}^{2}}}}
 \right) a_{{2}}b_{{0}}b_{{2}}+2\,\arctan \left( {\frac {2\,zb_{{2}}+b
_{{1}}}{\sqrt {4\,b_{{0}}b_{{2}}-{b_{{1}}}^{2}}}} \right) {b_{{1}}}^{2
}a_{{2}}
\nonumber \\
-4\,\arctan \left( {\frac {b_{{1}}}{\sqrt {4\,b_{{0}}b_{{2}}-{
b_{{1}}}^{2}}}} \right) a_{{0}}{b_{{2}}}^{2}+2\,\arctan \left( {\frac
{b_{{1}}}{\sqrt {4\,b_{{0}}b_{{2}}-{b_{{1}}}^{2}}}} \right) b_{{1}}a_{
{1}}b_{{2}}
\nonumber \\
+4\,\arctan \left( {\frac {b_{{1}}}{\sqrt {4\,b_{{0}}b_{{2}
}-{b_{{1}}}^{2}}}} \right) a_{{2}}b_{{0}}b_{{2}}-2\,\arctan \left( {
\frac {b_{{1}}}{\sqrt {4\,b_{{0}}b_{{2}}-{b_{{1}}}^{2}}}} \right) {b_{
{1}}}^{2}a_{{2}}
\nonumber
\end{eqnarray}
Figure \ref{errore_pade} reports the percentage
error, see formula (\ref{error100}),  
for $(m-M)_{2,2}$ as function of the redshift
until  the value 
of $1\%$ is reached at  $z \approx 6$. 
For $z > 6$ the Pad\'e approximant  of the
distance modulus does not converge to the
numerical distance modulus.

More details can be found in \cite{Zaninetti2016a}.

\subsection{ Dynamical dark energy or wCDM}

In the dynamical dark
energy  cosmology (wCDM),
firstly introduced by   \cite{Turner1997},
the {\it Hubble distance}   is    
\begin{equation}
D_H(z;\om,w,\ode) =
 \frac {1}{
\sqrt { 
\left( 1+z \right) ^{3}{\it \om}+{\it \ode}\, \left( 1+z
 \right) ^{3+3\,w}
}
 }
\quad ,
\label{dhwcdm}  
\end{equation}
where  $w$ is the equation of state  here considered  constant,
see equation (3.4) in  \cite{Tripathi2017} or
equation(18) in \cite{Wei2015} for the luminosity distance.
Here we considered $w$ to be constant but  also  the case 
of $w$ as function of $z$ can be   considered, see equation (19)
in \cite{Wei2015}.
In the above cosmology the cosmological constant is absent.
In flat cosmology
\begin{equation}
\om +\ode=1
\quad ,
\end{equation}
and the  {\it Hubble distance} becomes
\begin{equation}
D_H(z;\om,w) =
 \frac {1}
{
\sqrt { \left( 1+z \right) ^{3}{\it \om}+ \left( 1-{\it \om} \right) 
 \left( 1+z \right) ^{3+3\,w}}
 }
\quad .
\end{equation}
The indefinite integral  in the  variable  $z$ of the 
above  {\it Hubble distance}, 
$Iz \equiv \frac{D_C}{D_H}$,
is 
\begin{equation}
Iz(z;\om,w) =
\int D_H(z;\om,w)  dz 
\quad ,
\label{integralez_dm}
\end{equation}
where the new symbol  $Iz$ 
underline  the mathematical 
operation of integration.  
In order to solve for the the indefinite integral we  
perform a change of variable
$1+z=t^{1/3}$
\begin{equation}
Iz(t;\om,w) =
\frac{1}{3} \int \!{\frac {1}{\sqrt {-t \left(  \left( -1+{\it \om} \right)
{t}^{w}
-{\it \om} \right) }{t}^{2/3}}}\,{\rm d}t
\quad .
\label{integralet}
\end{equation}
The indefinite integral is
\begin{equation}
Iz(t;\om,w) =
\frac
{
-2\,
{\mbox{$_2$F$_1$}(\frac{1}{2},-\frac{1}{6}\,{w}^{-1};\,1-\frac{1}{6}\,{w}^{-1};\,-{\frac {{t}^{w}
- \left( 1-{\it \om} \right) }{{\it \om}}})}
}
{
\sqrt {{\it \om}}\sqrt [6]{t}
}
\quad ,
\label{hubbleintegral}
\end{equation}
where ${\2F1(a,b;\,c;\,z)}$ is the
regularized hypergeometric
function,
see  
\cite{Abramowitz1965,Seggern1992,Thompson1997,Gradshteyn2007,NIST2010}.
We now return to the  variable $z$, the redshift.
Then the indefinite integral becomes
\begin{eqnarray}
Iz(z;\om,w) = \nonumber \\
\frac
{
-2\,
{\mbox{$_2$F$_1$}(\frac{1}{2},-\frac{1}{6}\,{w}^{-1};\,1-\frac{1}{6}\,{w}^{-1};\,-{\frac { \left(
- {z}^{3}+3\,{z}^{2}+3\,z+1 \right) ^{w} \left( 1-{\it \om} \right) }{{\it
- \om}}})}
}
{
\sqrt {{\it \om}}\sqrt [6]{{z}^{3}+3\,{z}^{2}+3\,z+1}
}
\quad .
\end{eqnarray}
We denote by $F(z;\om,w)$ the definite integral
\begin{equation}
F(z;\om,w)  = Iz(z=z;\om,w)  - Iz(z=0;\om,w)
\quad .
\label{definitefz}
\end{equation} 
The luminosity distance, $\dl$,
for wCDM cosmology in the case of the
analytical solution  is  
\begin{equation}
  \dl(z;c,H_0,\om,w) = \frac{c}{H_0} (1+z) F(z;\om,w)
\quad ,
\end{equation}
where $F(z;\om,w)$  is  given by  
equation (\ref{definitefz})
and
the distance modulus 
is
\begin{equation}
(m-M) =25 +5 \log_{10}\bigg ( \dl(z;c,H_0,\om,w)     \bigg)
\quad  .
\label{distmod_dark}
\end{equation}
More details can be found in \cite{Zaninetti2019c}.

\subsection{The Cardassian cosmology}

In flat Cardassian cosmology \cite{Freese2002,Freese2003} 
the {\it Hubble distance} 
is 
\begin{equation}
D_H(z;\om,w,n) =
{\frac {1}{\sqrt { \left( 1+z \right) ^{3}{\it \om}+ \left( 1-{\it 
\om} \right)  \left( 1+z \right) ^{3\,n}}}}
\quad ,
\end{equation} 
where $n$ is a variable parameter, and 
$n=0$ means the $\Lambda$CDM cosmology,
see equation (17) in \cite{Wei2015}.
The above equation can  also be obtained 
inserting $n=1+w$ in equation (\ref{dhwcdm}).
Despite  of this fact  the FORTRAN code which derives 
the cosmological  parameters produces a small difference
in the results because the variables are evaluated
in a different way.
The indefinite integral  in the  variable  $z$ of the 
above  {\it Hubble distance}, $Iz$,  is 
\begin{equation}
Iz(z;\om,n) =
\int D_H(z;\om,n)  dz 
\quad .
\label{integralezcard}
\end{equation}
In order to obtain the indefinite integral we  
perform a change of variable
$1+z=t^{1/3}$
\begin{equation}
Iz(t;\om,n) =\frac{1}{3}
\int \!{\frac {1}{\sqrt {-{t}^{n}{\it \om}+{\it \om}\,t+{t}^{n}}{t}^{2/3
}}}\,{\rm d}t
\quad .
\label{integraletcard}
\end{equation}
The indefinite integral is
\begin{equation}
Iz(t;\om,n) =
\frac
{
-2\,
{\mbox{$_2$F$_1$}\Big(1/2,- \left( 6\,n-6 \right) ^{-1};\,{\frac
{6\,n-7}{6\,n-6}};\,{\frac {{t}^{n-1} \left( {\it \om}-1 \right) }{{\it
\om}}}\Big)}
}
{
\sqrt {{\it \om}}\sqrt [6]{t}
}
\quad ,
\label{hubbleintegralcard}
\end{equation}
where ${\2F1(a,b;\,c;\,z)}$ is the
regularized hypergeometric
function.
We now return to the original variable $z$
and the indefinite integral is
\begin{eqnarray}
Iz(z;\om,n) = \nonumber \\
\frac
{
-2\,
{\mbox{$_2$F$_1$}\Big(1/2,- \left( 6\,n-6 \right) ^{-1};\,{\frac
{6\,n-7}{6\,n-6}};
\,{\frac { \left(  \left( 1+z \right) ^{3} \right)^{n-1} 
\left( {\it \om} -1 \right) }{{\it \om}}}\Big)}
}
{
\sqrt {{\it \om}}\sqrt [6]{ \left( 1+z \right) ^{3}}
}
\quad .
\label{icardz}
\end{eqnarray}
We denote by $F_c(z;\om,n)$ the definite integral
\begin{equation}
F_c(z;\om,n)  = Iz(z=z;\om,n)  - Iz(z=0;\om,n)
\quad .
\label{definitefzcard}
\end{equation} 
In the case of the Cardassian cosmology,
the luminosity distance is 
\begin{equation}
  \dl(z;c,H_0,\om,n) = \frac{c}{H_0} (1+z) F_c(z;\om,n)
\quad ,
\end{equation}
where $F_c(z;\om,n)$  is  given by  
equation (\ref{definitefzcard})
and the distance modulus is
\begin{equation}
(m-M) =25 +5 \log_{10}\bigg ( \dl(z;c,H_0,\om,n)     \bigg)
\quad .
\label{distmod_cardassian}
\end{equation}
In the flat Cardassian cosmology, 
there are three parameters: $H_0,\om$ and $n$.
More details can be found in \cite{Zaninetti2019c}.

\subsection{The flat cosmology}

The starting point is
equation (1) for the luminosity 
distance in \cite{Baes2017} 
\begin{equation}
  \dl(z;c,H_0,\om) = \frac{c(1+z)}{H_0} \int_{0}^{z}\!{\frac {1}{\sqrt {{\it
\om}\, \left( 1+t \right) ^{3}
+1-{\it \om}}}}\,{\rm d}t
\quad ,
\end{equation}
where  the variable of integration, $t$, 
denotes the redshift.

A first 
change in the parameter $\om$ introduces
\begin{equation}
s=\sqrt [3]{{\frac {1-{\it \om}}{{\it \om}}}}
\quad 
\label{sdef}  
\end{equation}
and the luminosity distance becomes
\begin{equation}
  \dl(z;c,H_0,s) =\frac{1}{H_0}  
c \left( 1+z \right) \int_{0}^{z}\!{\frac {1}{\sqrt {{\frac { \left( 1
+t \right) ^{3}}{{s}^{3}+1}}+1- \left( {s}^{3}+1 \right) ^{-1}}}}
\,{\rm d}t
\quad .
\end{equation}
The following change of variable, $t=\frac{s-u}{u}$, is performed 
for the luminosity distance, which becomes 
\begin{eqnarray}
  \dl(z;c,H_0,s) =
\nonumber  \\
-\frac{c}{{\it H_0}\,{s}^{2}}  
 \left( 1+z \right)  \left( {s}^{3}+1 \right) \int_{s}^{{\frac {s}{1
+z}}}\!{\frac {u}{{u}^{3}+1}\sqrt {{\frac {{s}^{3} \left( {u}^{3}+1
 \right) }{{u}^{3} \left( {s}^{3}+1 \right) }}}}\,{\rm d}u
\quad .
\end{eqnarray}
The  integral
for the luminosity distance is
\begin{eqnarray}
  \dl(z;c,H_0,s) =
-1/3\,{\frac {c  ( 1+z   ) {3}^{3/4}\sqrt {{s}^{3}+1}}{\sqrt {
s}{\it H_0}}
} \times \nonumber \\
{
\Bigg  ( {\it F}  ( 2\,{\frac {\sqrt {s  ( s
+1+z   ) }\sqrt [4]{3}}{s\sqrt {3}+s+z+1}},1/4\,\sqrt {2}\sqrt {3}
+1/4\,\sqrt {2}   ) 
}
\nonumber \\
{
-{\it F}  ( 2\,{\frac {\sqrt [4]{3
}\sqrt {s  ( s+1   ) }}{s+1+s\sqrt {3}}},1/4\,\sqrt {2}\sqrt {
3}+1/4\,\sqrt {2}   )   \Bigg ) }
\quad ,
\label{distlumflatnew}
\end{eqnarray}
where $s$ is given by Eq. (\ref{sdef}) and 
$F\left(\phi,k\right)$
is Legendre’s incomplete elliptic integral of the first kind,
\begin{equation}
F\left(\phi,k\right)=
\int_{0}^{\sin\phi}\frac{\mathrm{d}t}{\sqrt{1-t^{2}}\sqrt{1-k^{2}t^{2}}}
\quad ,
\end{equation}
see  \cite{NIST2010}.
The distance modulus is
\begin{equation}
(m-M) =25 +5 \log_{10}\bigg ( \dl(z; c,H_0,s) \bigg)
\quad ,
\label{distmod_elliptic_new}
\end{equation}
and therefore 
\begin{eqnarray}
(m-M)= 25
\nonumber \\
+5\,{\frac {1}{\ln  \left( 10 \right) }\ln  \left( -\frac{1}{3}\,{\frac {c
 \left( 1+z \right) {3}^{3/4} \left( {\it F_1}-{\it F_2} \right) 
\sqrt {{s}^{3}+1}}{\sqrt {s}{\it H_0}}} \right) }
\quad ,
\label{distmod_flatnew}
\end{eqnarray}
where 
\begin{equation}
F_1=
{\it F} \left( 2\,{\frac {\sqrt {s \left( s+1+z \right) }
\sqrt [4]{3}}{s\sqrt {3}+s+z+1}},1/4\,\sqrt {2}\sqrt {3}+1/4\,\sqrt {2
} \right)
\end{equation}
and
\begin{equation}
F_2=
{\it F} \left( 2\,{\frac {\sqrt [4]{3}\sqrt {s \left( s+1
 \right) }}{s+1+s\sqrt {3}}},1/4\,\sqrt {2}\sqrt {3}+1/4\,\sqrt {2}
 \right) 
\quad , 
\end{equation}
with $s$ as defined  by Eq. (\ref{sdef}).
More details can be found in \cite{Zaninetti2019b}.

\subsection{$\phi$CDM cosmology}

\begin{figure}
\includegraphics[width=6cm]{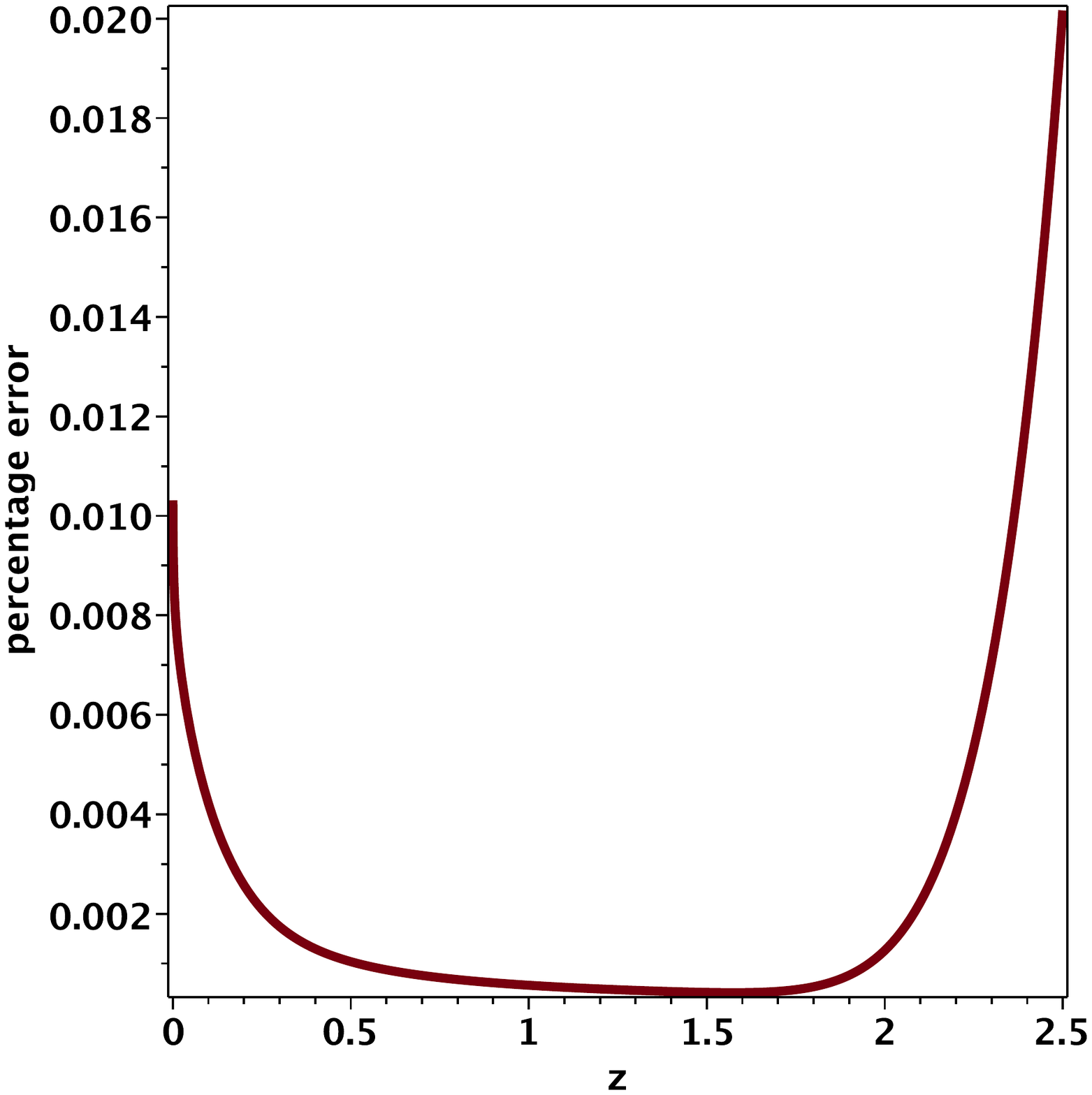}
\caption{
Percentage error  of $(m-M)_{7}$ in respect
to the numerical value
with  data as in Table \ref {chi2_union21}. 
}
\label{phipercent}
\end{figure}

The inflationary universe  has  been introduced by
\cite{Starobinsky1980,Guth1981,Ratra1988}
and  the term "quintessence" in a title of a paper 
appeared in \cite{Steinhardt1998}.
At the moment of  writing  given a  scalar field, 
$\phi$, and the connected  self-interacting potential, 
$V(\phi)$,  ten different  quintessence models 
are suggested by \cite{Avsajanishvili2018}.
Here we start from equation (12) in \cite{Mamon2017}
where $E(z)$, the `Hubble function', is
\begin{equation}
E(z;\zerom,\fizero\alpha\beta)=\sqrt { \left( 1+z \right) ^{3}{\it \zerom}+{\it \fizero}\, \left( 
1+z \right) ^{\alpha}{{\rm e}^{\beta\,z}}}
\quad ,
\end{equation}
where 
$\zerom=\frac{\rho_{m0}}{3 H_0^2}$ is the adimensional present 
density of matter,
$\fizero=\frac{\rho_{\phi\,0}}{3 H_0^2}$ is the present adimensional 
density of the scalar field, 
$H_0$ is the present value of the Hubble constant,  
$\rho_{m0}$ is the present density of matter,
$\rho_{\phi\,0}$ is the present density of the scalar field,
$\alpha$ and $\beta$ are two parameters 
which allow to match theory and observations.
In absence of curvature we have
\begin{equation}
\zerom + \fizero=1
\quad ,
\end{equation}
and
therefore 
\begin{equation}
E(z;\zerom,\alpha,\beta)=
\sqrt { \left( 1+z \right) ^{3}{\it \zerom}+ \left( 1-{\it \zerom}
 \right)  \left( 1+z \right) ^{\alpha}{{\rm e}^{\beta\,z}}}
\quad .
\end{equation}
The luminosity 
distance is  
\begin{equation}
  \dl(z;c,H_0,\zerom,\alpha,\beta) = \frac{c(1+z)}{H_0} \int_{0}^{z}\!
\frac{1}{E(t;\zerom,\alpha,\beta)} \,dt
\quad ,
\label{dlphi}
\end{equation}
where  the variable of integration, $t$, 
denotes the redshift.
At the moment of writing there is not  an analytical
solution for the above integral
and therefore we implement a numerical solution, 
$\dlnum(z;c,H_0,\zerom,\alpha,\beta)$.
The distance modulus is
\begin{equation}
(m-M) =25 +5 \log_{10}\bigg ( \dlnum(z;c,H_0,\zerom,\alpha,\beta) \bigg)
\quad .
\label{distmod_num_phi}
\end{equation}
An approximate value of the above integral (\ref{dlphi})
is obtained with a Taylor expansion of the integrand  
about $z=1$ of order seven  denoted by 
$\dlsette(z;c,H_0,\zerom,\alpha,\beta)$.
We report  the  numerical expression with cosmological 
parameters as in Table \ref{chi2_union21} 
relative to  the Union 2.1 compilation:
\begin{eqnarray}
\dlsette(z)=   \nonumber  \\
 4282.7\, \left( 1+z \right)  \Big (  0.91287\,z-
 0.16562\,{z}^{2}+ 0.039001\, \left( z-1 \right) ^{3}
\nonumber \\
-
 0.003084\, \left( z-1 \right) ^{4}
- 0.0036858\, \left( z-1
 \right) ^{5}+ 0.0028217\, \left( z-1 \right) ^{6}
\nonumber  \\
-
 0.00115816\, \left( z-1 \right) ^{7}+ 0.03442 \Big ) 
\quad .
\end{eqnarray}
The approximate distance modulus is
\begin{equation}
(m-M)_7 =25 +5 \log_{10}\bigg ( \dlsette(z;c,H_0,\zerom,\alpha,\beta) \bigg)
\quad ,
\label{distmod_num_phi_app}
\end{equation}
which for  the Union 2.1 compilation
has the following numerical expression
\begin{eqnarray}
(m-M)_7
=
25
\nonumber  \\
+\frac{5}{\ln    ( 10   ) }  \Bigg (  \ln \Big   (  4282.7    ( 1+z   )    ( 
 0.91287 z- 0.16562 {z}^{2}+ 0.039001    ( z-1
   ) ^{3}
\nonumber \\
- 0.0030847    ( z-1   ) ^{4}-
 0.0036858    ( z-1   ) ^{5}+ 0.0028217    ( z-1
   ) ^{6}
\nonumber \\
- 0.0011581    ( z-1   ) ^{7}
+ 0.03442
   )   \Big )  \Bigg )
\, .
\end{eqnarray}
Figure \ref{phipercent} reports the percentage
error, see formula (\ref{error100}),  
for $(m-M)_{7}$ as function of the redshift
until  the value 
of $0.02\%$ is reached at  $z \approx 2.5$. 

\subsection{The Einstein--De Sitter cosmology}

In the Einstein--De Sitter model
the luminosity distance,$D_L$, after \cite{Einstein1932,Krisciunas1993},
is
\begin{equation}
D_L =2\,{\frac {c \left( 1+z-\sqrt {z+1} \right) }{H_{{0}}}}
\quad ,
\end{equation}
and the distance modulus for the Einstein--De Sitter model is
\begin{equation}
m-M =25+5\,{\frac {1}{\ln  \left( 10 \right) }\ln  \left( 2\,{\frac {c
 \left( 1+z-\sqrt {z+1} \right) }{H_{{0}}}} \right) }
\quad .
\label{distmod_eds}
\end{equation}
There is one free parameter in the Einstein--De Sitter model: $H_0$.
The Einstein--De Sitter model has  been  recently  improved 
by  \cite{Gupta2018}, splitting the analysis in
two:  the Einstein--De Sitter flat, only-matter universe,
referred to as EdesNa, and a flat, only-matter, including the Mach effect
universe,  referred to as EDSM.
We limit ourselves  to the EdesNA model
and we start from  equation (37) of 
\cite{Gupta2018}
\begin{equation}
m-M =5\,{\frac {\ln  \left( 5/3\,R_{{0}} \left( 1+z \right) {\it I_G(z)}
 \right) }{\ln  \left( 10 \right) }}+25
\quad ,
\label{distmod_edsna}
\end{equation}
where
\begin{equation}
R_0 =\frac{c}{H_0}
\quad ,
\end{equation}
and
\begin{equation}
I_G(z)=\int_0^z  \frac{1}{1+{\frac {2}{3} \left( 1+x \right) ^{{\frac{3}{2}}}}}
dx
\label{integral_gupta}
\quad .
\end{equation}
Evaluating the
integral yields
\begin{eqnarray}
I_G(z)
=
-\sqrt [6]{3} \Bigg  ( \arctan   ( {\frac {2\,\sqrt [6]{3}\sqrt [3]{2
}}{3}}-{\frac {\sqrt {3}}{3}}   ) 
\nonumber \\
-\arctan   ( {\frac {2\,
\sqrt [6]{3}\sqrt [3]{2}}{3}\sqrt [3]{   ( 1+z   ) ^{{\frac{3}{
2}}}}}-{\frac {\sqrt {3}}{3}}   )   \Bigg ) \sqrt [3]{2}-{\frac {{
12}^{{\frac{2}{3}}}}{12}  \Bigg ( \ln    ( -\sqrt [3]{2}\sqrt [3]{3}+
{2}^{{\frac{2}{3}}}+{3}^{{\frac{2}{3}}}   ) 
}
\nonumber  \\
{
-2\,\ln    ( \sqrt 
[3]{2}+\sqrt [3]{3}   ) +2\,\ln    ( \sqrt [3]{2}{3}^{2/3}
\sqrt [3]{   ( 1+z   ) ^{3/2}}+3   ) 
}
\nonumber  \\
{
-\ln    ( {2}^{{
\frac{2}{3}}}\sqrt [3]{3}   (    ( 1+z   ) ^{{\frac{3}{2}}}
   ) ^{{\frac{2}{3}}}-\sqrt [3]{2}{3}^{{\frac{2}{3}}}\sqrt [3]{
   ( 1+z   ) ^{{\frac{3}{2}}}}+3   ) -\ln    ( 3
   )   \Bigg ) }
\quad .
\end{eqnarray}
The integrand of (\ref{integral_gupta})
can be approximated  with 
a Pad\'e approximant with $p=2,q=2$,
\begin{equation}
I_{G22}(z)=\int_0^z  
\frac
{
-3\,{x}^{2}+36\,x+144
}
{
67\,{x}^{2}+204\,x+240
}
dx
\quad ,
\end{equation}
and therefore  we have the approximate integral
\begin{eqnarray}
I_{G22}(z)
=
-{\frac {3\,z}{67}}+{\frac {1512\,\ln  \left( 67\,{z}^{2}+204\,z+240
 \right) }{4489}}
\nonumber \\
+{\frac {64368\,\sqrt {1419}}{2123297}\arctan \left( 
{\frac { \left( 134\,z+204 \right) \sqrt {1419}}{5676}} \right) }-{
\frac {1512\,\ln  \left( 240 \right) }{4489}}
\nonumber \\
-{\frac {64368\,\sqrt {
1419}\arctan \left( {\frac {17\,\sqrt {1419}}{473}} \right) }{2123297}
}
\quad  ,
\end{eqnarray}
which generates the following approximate distance modulus
\begin{equation}
(m-M)_{22} =5\,{\frac {\ln  \left( 5/3\,R_{{0}} \left( 1+z \right) {\it
I_{G22}(z)}
 \right) }{\ln  \left( 10 \right) }}+25
\quad .
\label{distmod_edsna_app}
\end{equation}
The percent error between 
the approximate distance modulus as
given by equation (\ref{distmod_edsna_app})  
and the 
the exact  distance modulus as
given by equation (\ref{distmod_edsna})
is $\approx \,0.03 \%$  when $z=4$ and $H_0=69.1$.  

\subsection{Simple GR cosmology}

In the framework of GR,
the  received flux, $f$, is
\begin{equation}
f=\frac{L}{4\,\pi D_L^2}
\quad  ,
\end{equation}
where $D_L$ is the luminosity distance, which depends on the
cosmological model adopted, see Eq.~(7.21) in
\cite{Ryden2003} or Eq.~(5.235) in \cite{Lang2013}.

The distance modulus in the simple GR cosmology  is
\begin{equation}
m-M =
43.17-{\frac {1}{\ln  \left( 10 \right) }\ln  \left( {\frac {{\it H_0}
}{70}} \right) }+5\,{\frac {\ln  \left( z \right) }{\ln  \left( 10
 \right) }}+ 1.086\, \left( 1-{\it q_0} \right) z
\quad  ,
\label{distmod_simplegr}
\end{equation}
see  Eq.~(7.52) in  \cite{Ryden2003}.
There are two free parameters in the simple GR cosmology:
$H_0$ and $q_0$.

\subsection{Flat expanding universe}

This model is based on the standard definition
of luminosity in the flat expanding universe.
The luminosity distance, $r_{L}^{\prime}$, is
\begin{equation}
r_L^{\prime} =\frac{c}{H_0} z
\quad ,
\end{equation}
and the distance modulus
is
\begin{equation}
m-M = = -5 \log_{10}  +5 \log_{10}  r_L^{\prime} +2.5 \log(1+z)
\quad  ,
\label{distmod_heymann}
\end{equation}
see formulae (13) and (14) in \cite{Heymann2013}.
There is one free parameter in the flat expanding model,
$H_0$.

\subsection{The Milne universe in SR}

In the Milne model, which is developed in the framework of SR,
the luminosity distance,
after \cite{Milne1933,2005Chodorowski,Adamek2014}, is
\begin{equation}
D_L ={\frac {c \left( z+\frac{1}{2}\,{z}^{2} \right) }{H_{{0}}}}
\quad ,
\end{equation}
and the distance modulus for the Milne model is
\begin{equation}
m-M =
25+5\,{\frac {1}{\ln  \left( 10 \right) }\ln  \left( {\frac {c \left(
z+\frac{1}{2}\,{z}^{2} \right) }{H_{{0}}}} \right) }
\quad .
\label{distmod_milne}
\end{equation}
There is one free parameter in the Milne model: $H_0$.

\subsection{Plasma cosmology}

In an Euclidean static framework
from among many possible absorption mechanisms, we have selected
a plasma effect which produces the
following relation for the distance $d$
\begin{equation}
d= \frac{c}{H_0} \ln (1+z)
\quad ,
\label{nonlzdari}
\end{equation}
where the distance expressed  in  lower case 
underline the difference with the relativistic case,
see Eq. (50) in \cite {Brynjolfsson2004}.

In the presence of plasma absorption,
the observed flux is
\begin{equation}
f = \frac{{L  \cdot \exp \left( { - b  d  -
H_0 d - 2 H_0 d} \right)}}{{4 \pi d^2 }}
\quad ,
\end{equation}
where the factor $\exp \left( { - b  d } \right)$
is due to  galactic and host galactic extinctions,
$-H_0 d$ is the reduction due to the plasma in the IGM and
$- 2 H_0 d$  is the reduction due to the Compton scattering,
see the formula before Eq.~(51) in \cite {Brynjolfsson2004}.
The resulting distance modulus in the plasma mechanism
is
\begin{eqnarray}
m-M =
5\,{\frac {\ln  \left( \ln  \left( z+1 \right)  \right) }{\ln  \left(
10 \right) }}+\frac{15}{2}\,{\frac {\ln  \left( z+1 \right) }{\ln  \left( 10
 \right) }}
\nonumber \\
+5\,{\frac {1}{\ln  \left( 10 \right) }\ln  \left( {\frac {
c}{H_{{0}}}} \right) }+25+ 1.086\,b
\quad ,
\label{distmod_plasma}
\end{eqnarray}
see Eq.~(7) in \cite{Brynjolfsson2006}.
There is one free parameter in the plasma cosmology: $H_0$ when $b=0$.
A detailed analysis of this and other
physical mechanisms which produce
the observed redshift can be
found in \cite{Marmet2018}.

\subsection{Modified tired light}

\label{subsecplasma}
In an Euclidean static  universe,
the concept of modified tired light (MTL)  was introduced in
Section 2.2 of \cite{Zaninetti2015a}.
The distance in the MTL is
\begin{equation}
d= \frac{c}{H_0} \ln (1+z)
\quad ,
\label{nonlzd}
\end{equation}
where  the distance expressed  in  lower case 
underline the difference with the relativistic case.
The  distance modulus in  MTL
is
\begin{equation}
m-M = \frac{5}{2}\,{\frac {\beta\,\ln  \left( z+1 \right) }{\ln  \left( 10 \right) }
}+5\,{\frac {1}{\ln  \left( 10 \right) }\ln  \left( {\frac {\ln
 \left( z+1 \right) c}{H_{{0}}}} \right) }+25
\quad ,
\label{distmod_tired}
\end{equation}
where $\beta$ is a parameter lying  between 1 and 3
which allows matching theory with observations.
There are two free  parameters in MTL: $H_0$ and $\beta$.

\section{Astrophysical results}
\label{section_results}
We first review the statistics 
involved and then we process the 12$\times$4 cosmological cases.

\subsection{The adopted statistics}

In the case of the distance modulus, 
the  merit function $\chi^2$ is
\begin{equation}
\chi^2  = \sum_{i=1}^N \biggr [\frac{(m-M)_i - (m-M)(z_i)_{th}}{\sigma_i}\biggl] ^2
\quad ,
\label{chisquare}
\end{equation}
where $N$ is the number of SNs, 
$(m-M)_i$ is the observed distance modulus
evaluated at a redshift of $z_i$,
$\sigma_i$ is the error in the observed distance
modulus evaluated at $z_i$,
and
$(m-M)(z_i)_{th}$ is the theoretical distance modulus 
evaluated at $z_i$, see formula (15.5.5) in \cite{press}.
The reduced  merit function $\chi_{red}^2$
is  
\begin{equation}
\chi_{red}^2 = \chi^2/NF
\quad,
\label{chisquarereduced}
\end{equation}
where $NF=N-k$ is the number of degrees  of freedom,
       $N$     is the number of SNs,
and    $k$     is the number of free parameters.
Another useful statistical parameter is  the associated $Q$-value,
which has   to be understood as the
 maximum probability of obtaining a better fitting,
 see formula (15.2.12) in \cite {press}:
\begin{equation}
Q=1- GAMMQ (\frac{N-k}{2},\frac{\chi^2}{2} )
\quad ,
\end{equation}
where GAMMQ is a subroutine  for the incomplete gamma function.
The Akaike information criterion
(AIC), see \cite{Akaike1974},
is defined by
\begin{equation}
AIC  = 2k - 2  ln(L)
\quad,
\end{equation}
where $L$ is
the likelihood  function.
We assume  a Gaussian distribution for  the errors;
then  the likelihood  function
can be derived  from the $\chi^2$ statistic
$L \propto \exp (- \frac{\chi^2}{2} ) $
where  $\chi^2$ has been computed by
Eq.~(\ref{chisquare}),
see~\cite{Liddle2004}, \cite{Godlowski2005}.
Now the AIC becomes
\begin{equation}
AIC  = 2k + \chi^2
\quad .
\label{AIC}
\end{equation}
The    goodness of the approximation 
in evaluating a physical 
variable  $p$  is evaluated by the percentage 
error
$\delta$ 
\begin{equation}
\delta = \frac{\big | p - p_{approx} \big |}
{p} \times 100
\quad ,
\label{error100}
\end{equation}
where $p_{approx}$ is an approximation of $p$.

\subsection{The numerical techniques}

The parameters  of the twelve cosmologies here analyzed 
are found  minimizing  the
$\chi^2$   as given by equation (\ref{chisquare}).
We now report the adopted numerical techniques:
\begin{enumerate}
\item  In absence of an analytical solution 
       for the distance modulus we do
       k (the number of free parameters)  nested numerical
       loops for  the evaluation of the $\chi^2$.
       The  parameters which minimize the $\chi^2$
       are selected.
       This method allows to  find,  as an example,  
       the 
       parameters of the $\Lambda$CDM and $\phi$CDM   
       cosmologies.
\item  In presence  of an analytical solution,
       an approximate  Taylor     series
       and  
       a  Pad\'e approximant 
       for the distance modulus
       we derive  the parameters trough the
       Levenberg--Marquardt  method
       (subroutine MRQMIN in \cite{press})
       once  an analytical expression
       for the  derivatives  of the distance modulus
       with respect to the unknown  parameters is  provided.
       In absence of a human expression for the 
       derivatives we implement the numerical derivative.
       This method  was used  to evaluate the parameters 
       of  
       the MTL ,
       the  simple GR,
       the plasma,
       the Milne,
       the Einstein--De Sitter,
       the flat,
       the wCDM  and
       the Cardassian cosmologies.
\end{enumerate}

The  above  techniques 
allow to derive the cosmological parameters 
with unprecedented accuracy,
as an example  an error  of $0.1~\h0units$ 
can be associated with the Hubble constant.
The advantage to have approximate results , i.e. 
the Pad\'e approximant for the distance modulus  $(m-M)_{2,2}$
as given by equation (\ref{distmod_lcdm}),
is that we can evaluate  in an analytical
way  the first derivative  required by
the Levenberg--Marquardt  method  and the numerical 
integration is not necessary.

\subsection{The four compilations}

\begin{figure}
\includegraphics[width=6cm]{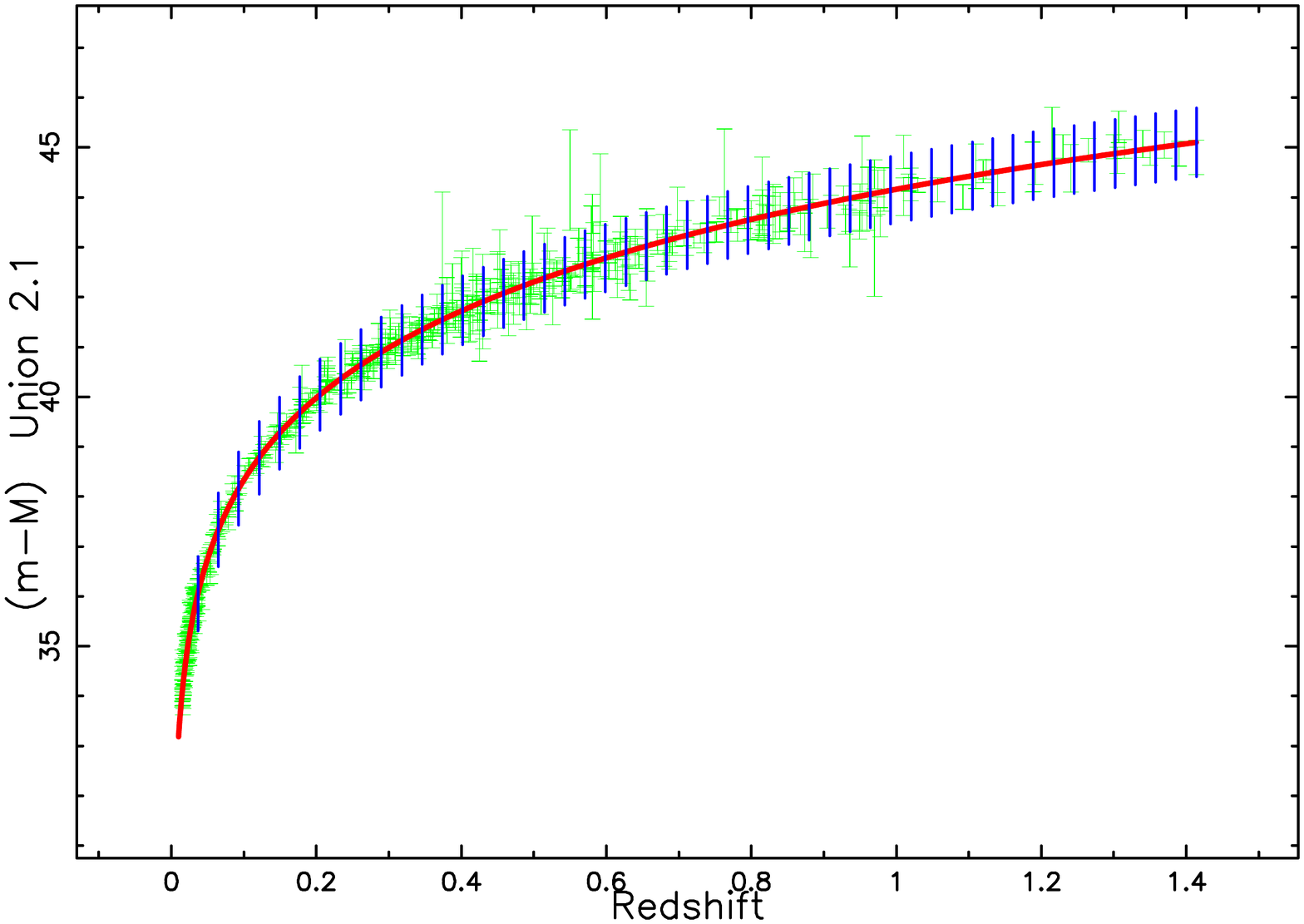}
\caption{
Hubble diagram for the  Union 2.1  compilation , green points
with error bar.
The solid red line represents the best fit
for the distance modulus  in 
$\Lambda$CDM  cosmology  
as represented by Eq.~(\ref{distmod_lcdm}).
The theoretical uncertainties are  represented through 
blue vertical lines  by applying the law of errors of
Gauss with the uncertainties and 
parameters as in the first line of Table 
\ref{chi2_union21}.
}
\label{union_distmrev}
\end{figure}

In order to avoid the  degeneracy in the Hubble constant-absolute magnitude 
plane we deal only with already calibrated distance modulus.
The first  astronomical test
we perform is on the 580 SNs  of  the Union 2.1 compilation, see \cite{Suzuki2012},
which is available at
\url{http://supernova.lbl.gov/Union/figures/SCPUnion2.1_mu_vs_z.txt}:
in this  compilation a calibrated distance versus redshift 
is provided.
The cosmological  parameters are reported
in  Table \ref{chi2_union21}
and Figure \ref{union_distmrev} reports the 
best fit in the $\Lambda$CDM cosmology.

\begin{table}[ht!]
\caption
{
Numerical values of  $\chi^2$, $\chi_{red}^2$, $Q$
and the AIC of the Hubble diagram for the Union 2.1 compilation:
$k$ stands for the number of parameters,
$H_0$ is expressed in $\h0units$; 580 SNs.
}
\label{chi2_union21}
\begin{center}
\resizebox{12cm}{!}
{
\begin{tabular}{|c|c|c|c|c|c|c|c|}
\hline
cosmology & Eq. &  $k$         &   parameters    & $\chi2$& $\chi_{red}^2$ &
$Q$  & AIC \\
\hline
$\Lambda$CDM & (\ref{distmod_lcdm}) &  3
& $H_0 = (69.56\pm 0.1)$ ; $\om=( 0.238\pm 0.01)$; $\ola=( 0.661 \pm0.01)$
& 562.59 &  0.975  & 0.658 & 569.39 \\
\hline
wCDM  
& (\ref{distmod_dark}) 
& 3 
& $H_0=(70.02\pm 0.35)$; 
$\om =(0.277\pm 0.025)$ 
; $w=  (-1.003\pm 0.05)$ 
&  562.21 &  0.974 & 0.662 & 568.21
\\
\hline
Cardassian 
& (\ref{distmod_cardassian})  
& 3
& $H_0=(70.15  \pm 0.38 )$; 
$\om =(0.305   \pm 0.019)$ 
; $n=  (-0.081 \pm 0.01 )$ 
&  562.35 &  0.974 & 0.661 
&  568.35
\\
\hline
flat &(\ref{distmod_flatnew}) &  2
& $H_0=69.77\pm 0.33$; $\om= 0.295\pm0.008$
& 562.55 &  0.9732  & 0.66 & 566.55 \\
\hline
$\phi$CDM   &(\ref{distmod_num_phi}) &  4
& $H_0=70\pm 0.1$; $\zerom= 0.28\pm0.02$; $\alpha=-0.08 \pm 0.2$; 
$\beta=0.05 \pm 0.02$
& 562.23 &  0.976  & 0.65 & 570.23 \\
\hline 
Einstein--De Sitter  & (\ref{distmod_eds}) & 1  &
$H_0=63.17\pm 0.2 $
& 1171.39 & 2.02& 2 $\,10^{-42}$  &  1173.39\\
\hline 
EdesNa  & (\ref{distmod_edsna}) & 1  &
$H_0=69.04\pm 0.22 $
& 569.46 & 0.98 &  0.603  &  571.46 \\
\hline 
simple~GR    & (\ref{distmod_simplegr}) & 2  
&$H_0=73.79\pm 0.024$, $q_0$=-0.1
& 689.34 &  1.194  & 9.5 $\,10^{-4}$ &  693.34\\
\hline
flat~expanding~model & (\ref{distmod_heymann}) & 1  &
$H_0=66.84\pm 0.22 $
&653  & 1.12 &  0.017  &  655 \\
\hline
Milne  & (\ref{distmod_milne}) & 1  &
$H_0=67.53 \pm 0.22 $
& 603.37 & 1.04 &  0.23 &  605.37\\
\hline
plasma       & (\ref{distmod_plasma}) & 1 & $H_0=74.2\pm 0.24 $
& 895.53 & 1.546   & 5.2 $\,10^{-16}$  & 897.5 \\
\hline
MTL  & (\ref{distmod_tired}) & 2 & $\beta$=2.37, $H_0=69.32 \pm 0.34 $  &
567.96 & 0.982 & 0.609 &571.9 \\
\hline
\end{tabular}
}
\end{center}
\end{table}

\begin{figure}
\includegraphics[width=6cm]{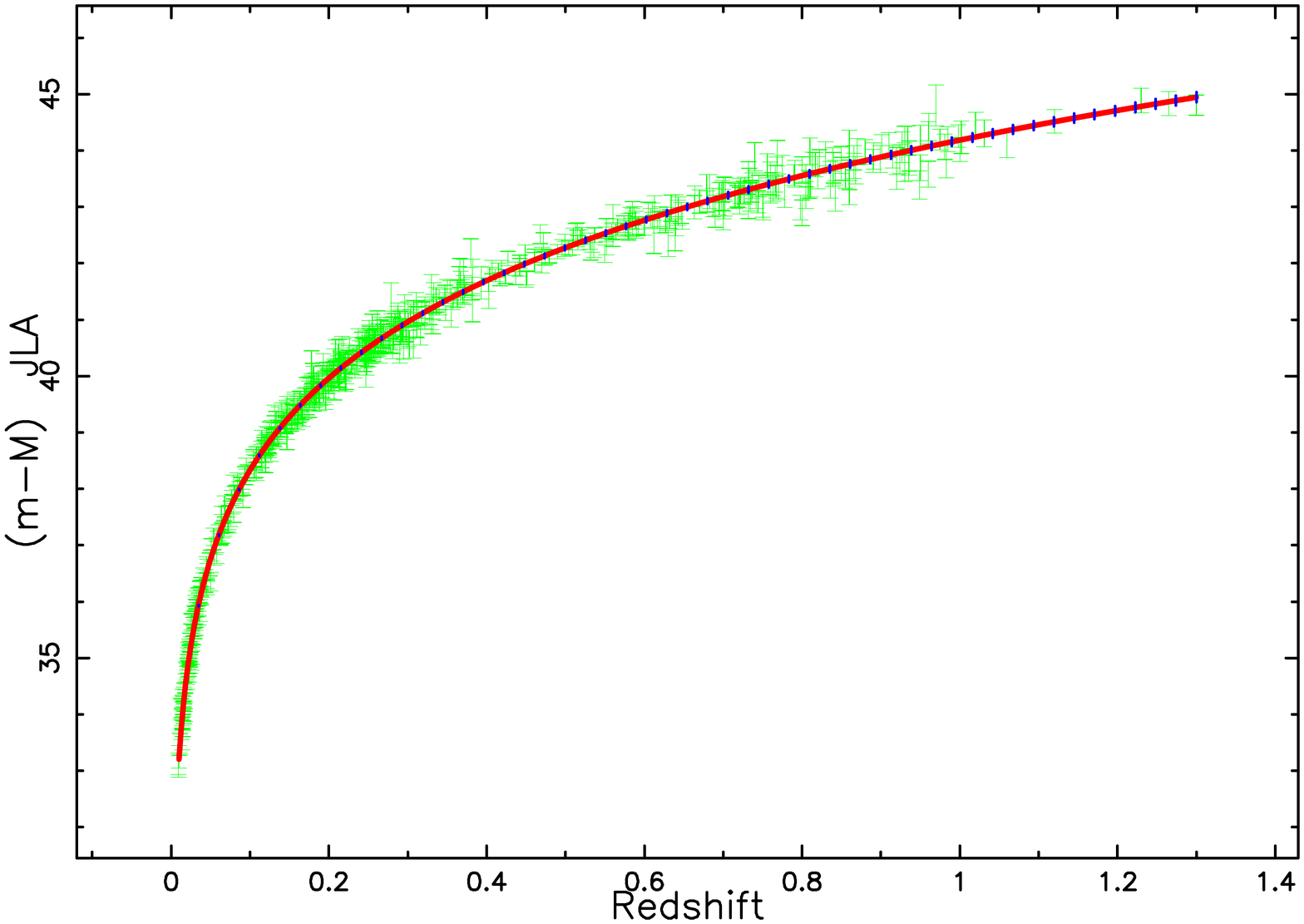}
\caption{
Hubble diagram for the  JLA  compilation, green points
with error bar.
The solid red line represents the best fit
for the distance modulus  in 
MTL  cosmology  
as represented by Eq.~(\ref{distmod_tired}).
The theoretical uncertainties are  represented through 
blue vertical lines.
}
\label{distmrev_jla}
\end{figure}

The second test we  perform 
is on the
the joint light-curve analysis (JLA),
which contains  740 SNs  \cite{Betoule2014} 
with data available on CDS at \url{http://cdsweb.u-strasbg.fr/}.
The above compilation consists of  SNe (type I-a)  for which
we have  a heliocentric redshift, $z$, apparent
magnitude $\mstar$ in the B band, error in $\mstar$, $\sigma_{\mstar}$,
parameter $X1$,   error in $X1$,
$\sigma_{X1}$,
parameter $C$, error in the parameter $C$, $\sigma_C$  and
$\log_{10} (M_{stellar})$.
The observed distance modulus  is defined by
Eq.~(4) in \cite{Betoule2014}
\begin{equation}
m-M =
-C\beta+{\it X1}\,\alpha-M_{{b}}+ \mstar
\quad.
\end{equation}
The  adopted parameters are
$\alpha=0.141$, $\beta=3.101$   and
\begin{equation}
M_{{b}} = \begin{cases}
-19.05  & \text{if } M_{stellar} <    10^{10} M_{\sun} \\
-19.12  & \text{if } M_{stellar} \geq 10^{10} M_{\sun}
 \end{cases}
\quad ,
\end{equation}
where $M_{\sun}$ is the mass of the sun,
see line 1  in Table 10 of
\cite{Betoule2014}.
The uncertainty  in the observed distance modulus,
$\sigma_{m-M}$,
is  found by implementing the error
propagation equation (often called the law of errors of Gauss) when
the covariant terms are neglected, see  equation (3.14)
in \cite{bevington2003},
\begin{equation}
\sigma_{m-M} =
\sqrt {{\alpha}^{2}{\sigma_{{{\it X1}}}}^{2}+{\beta}^{2}{\sigma_{{C}}}
^{2}+{\sigma_{{{\it \mstar}}}}^{2}}
\quad .
\end{equation}
The cosmological parameters with the JLA compilation
are reported in see Table \ref{chi2_jla}
and Figure \ref{distmrev_jla} reports the 
best fit in the MTL  cosmology.
\begin{table}[ht!]
\caption
{
Numerical values of  $\chi^2$, $\chi_{red}^2$, $Q$
and the AIC of the Hubble diagram for the JLA compilation,
$k$ stands for the number of parameters,
$H_0$ is expressed in $\h0units$; 740 SNs.
}
\label{chi2_jla}
\begin{center}
\resizebox{12cm}{!}
{
\begin{tabular}{|c|c|c|c|c|c|c|c|}
\hline
cosmology & Eq. &  $k$         &   parameters    & $\chi2$& $\chi_{red}^2$ &
$Q$  & AIC \\
\hline
$\Lambda$CDM & (\ref{distmod_lcdm}) &  3
& $H_0 = (70.71\pm 0.1)$; $\om=(0.238 \pm 0.01)$; $\ola=(0.621 \pm0.01)$
& 626.53 & 0.85  &  0.998 &  632.53 \\
\hline
wCDM  
& (\ref{distmod_dark}) 
& 3 
& $H_0=(69.38 \pm 0.31)$; 
$\om =(0.2  \pm  0.016)$ 
; $w=  (-0.8\pm   0.031)$ 
&  626.01 &  0.849 & 0.998 & 632.01
\\
\hline
Cardassian 
& (\ref{distmod_cardassian})  
& 3
& $H_0=(70.03  \pm 0.44 )$; 
$\om =(0.3     \pm 0.019)$ 
; $n=  (-0.055 \pm 0.004 )$ 
&  628.73 &  0.853 & 0.998 
&  634.73
\\
\hline
flat  &(\ref{distmod_flatnew}) &  2
& $H_0=69.65\pm 0.23$; $\om= 0.3\pm0.003$
& 627.91 &  0.85  & 0.998 & 631.91 \\
\hline
$\phi$CDM   &(\ref{distmod_num_phi}) &  4
& $H_0=69.6\pm 0.1$; $\zerom= 0.24\pm0.02$; $\alpha=0.31 \pm 0.2$; 
$\beta=0.03 \pm 0.02$
& 626.52 &  0.851  & 0.998 & 634.52 \\
\hline 
Einstein--De Sitter  & (\ref{distmod_eds}) & 1  &
$H_0= 62.57 \pm 0.17 $
& 1307.75 &1.76 &  $3.27\,10^{-34}$  &  1309.75\\
\hline 
EdesNa  & (\ref{distmod_edsna}) & 1  &
$H_0=68.91 \pm  0.19 $
& 630.46 & 0.853 &  0.998  &   632.46 \\
\hline 
simple~GR  & (\ref{distmod_simplegr}) & 2  &$H_0=73.79\pm 0.023$, $q_0$=-0.14
& 749.14 &   1.016 & 0.369 &  755.14\\
\hline
flat~expanding~model & (\ref{distmod_heymann}) & 1  &
$H_0=66.49 \pm 0.18 $
&717.3  & 0.97 &  0.709  &  719.3 \\
\hline
Milne  &(\ref{distmod_milne}) & 1  &
$H_0=67.19 \pm 0.18 $
& 656.11 & 0.887 &  0.986 &  658.11\\
\hline
plasma    & (\ref{distmod_plasma})& 1 & $H_0=74.45\pm 0.2$
& 1017.79 & 1.377   & 3.59 $\,10^{-11}$  & 1019.79 \\
\hline
MTL & (\ref{distmod_tired}) & 2 & $\beta$=2.36, $H_0=69.096 \pm 0.32$ &
626.27 & 0.848 & 0.998 &630.27\\
\hline
\end{tabular}
}
\end{center}
\end{table}

\begin{figure}
\includegraphics[width=6cm]{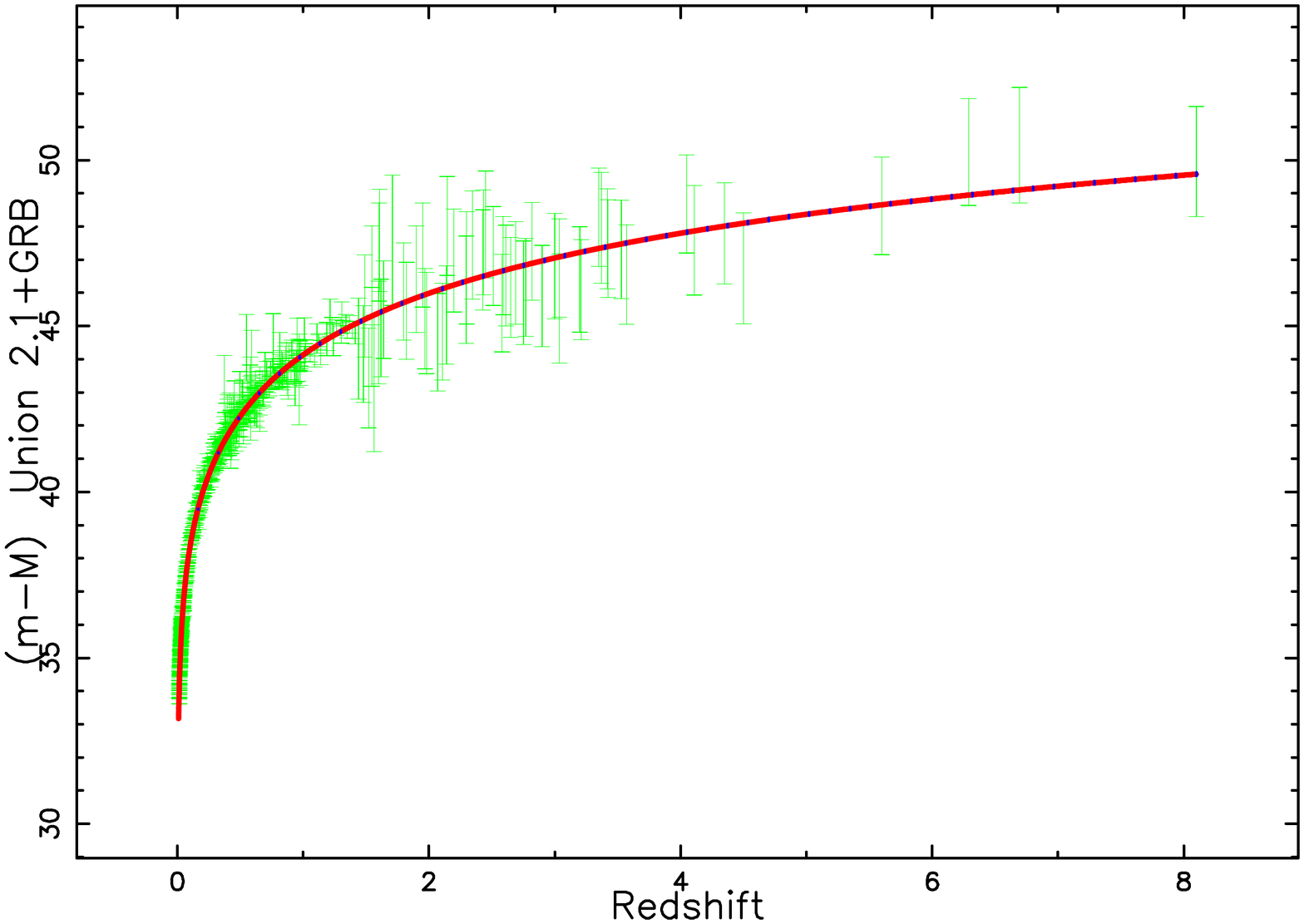}
\caption{
Hubble diagram for the  Union 2.1  compilation + 
the `Hymnium'  GRB sample, green points
with error bar.
The solid red line represents the best fit
for the distance modulus  in 
Cardassian  cosmology  
as represented by Eq.~(\ref{distmod_cardassian}).
The theoretical uncertainties are  represented through 
blue vertical lines.
}
\label{distmrev_grb_union}
\end{figure}

The third test is performed on the
Union 2.1  compilation (580 SNs) + 
the distance modulus for 59 
calibrated high-redshift GRBs, the  
so called  `Hymnium' sample of GRBs, 
which   allows to calibrate the distance
modulus in the high redshift up 
to $z \approx 8 $ \cite {Wei2010},
see Table \ref{chi2_union_grb}
and Figure \ref{distmrev_grb_union} for the 
best fit in the Cardassian  cosmology.

\begin{table}[ht!]
\caption
{
Numerical values of  $\chi^2$, $\chi_{red}^2$, $Q$
and the AIC of 
the Hubble diagram for 
the Union 2.1  compilation + 
the `Hymnium'  GRB sample,
$k$ stands for the number of parameters,
$H_0$ is expressed in 
$\h0units$; 580 SNs +59 GRBs.
}
\label{chi2_union_grb}
\begin{center}
\resizebox{12cm}{!}
{
\begin{tabular}{|c|c|c|c|c|c|c|c|}
\hline
cosmology & Eq. &  $k$         &   parameters    & $\chi2$& $\chi_{red}^2$ &
$Q$  & AIC \\
\hline
$\Lambda$CDM & (\ref{distmod_lcdm}) &  3
& $H_0 = (67.8  \pm 0.2)$ ; $\om=(0.259\pm0.02)$; $\ola=(0.691 \pm 0.02)$
&   586.04 & 0.921  & 0.922 &  592.04 \\
\hline
wCDM  
& (\ref{distmod_dark}) 
& 3 
& $H_0=(69.34 \pm 0.32)$; 
$\om =(0.2  \pm  0.016)$ 
; $w=  (-0.626\pm   0.015)$ 
&  592.1 &  0.93  & 0.892 & 598.1
\\
\hline
Cardassian 
& (\ref{distmod_cardassian})  
& 3
& $H_0=(70.1    \pm 0.42   )$; 
$\om =( 0.299   \pm 0.019  )$ 
; $n=  (-0.063  \pm  0.009 )$ 
&  585.43 &  0.92 & 0.924 
&  591.43
\\
\hline
flat  &(\ref{distmod_flatnew}) &  2
& $H_0=69.82\pm 0.24$; $\om=  0.295 \pm0.003$
&  585.74 &  0.919  & 0.927 & 589.74 \\
\hline
$\phi$CDM   &(\ref{distmod_num_phi}) &  4
& $H_0=70\pm 0.1$; $\zerom= 0.28\pm0.02$; $\alpha=-0.07 \pm 0.2$; 
$\beta=0.05 \pm 0.02$
& 585.41 &  0.922  & 0.92 & 593.41 \\
\hline 
Einstein--De Sitter  & (\ref{distmod_eds}) & 1  &
$H_0=  63.14 \pm    0.2 $
&  1205.2 & 1.88 & $3.58\,10^{-37}$  &  1205.21\\
\hline 
EdesNa  & (\ref{distmod_edsna}) & 1  &
$H_0=69.05 \pm  0.22 $
& 592.79 & 0.929 &  0.899  &    594.79 \\
\hline 
simple~GR  & (\ref{distmod_simplegr}) & 2  &$H_0=73.79\pm 0.023$,
$q_0$=-0.01
&  809.5 &   1.27 &  $3.85\,10^{-6}$ &  813.5\\
\hline
flat~expanding~model & (\ref{distmod_heymann}) & 1  &
$H_0= 66.851 \pm 0.22 $
& 676.36  & 1.06 & 0.141  & 678.36 \\
\hline
Milne  &(\ref{distmod_milne}) & 1  &
$H_0=67.55 \pm  0.22 $
&  634.27 & 0.994 &  0.534 &  636.27\\
\hline
plasma    & (\ref{distmod_plasma})& 1 & $H_0= 74.25 \pm 0.24$
&  951.16 & 1.49   & 9.39 $\,10^{-14}$  & 953.16  \\
\hline
MTL & (\ref{distmod_tired}) & 2 & $\beta$=2.35 , $H_0= 69.23 \pm 0.34$ &
594.69 & 0.933 & 0.883 & 598.69\\
\hline
\end{tabular}
}
\end{center}
\end{table}

\begin{figure}
\includegraphics[width=6cm]{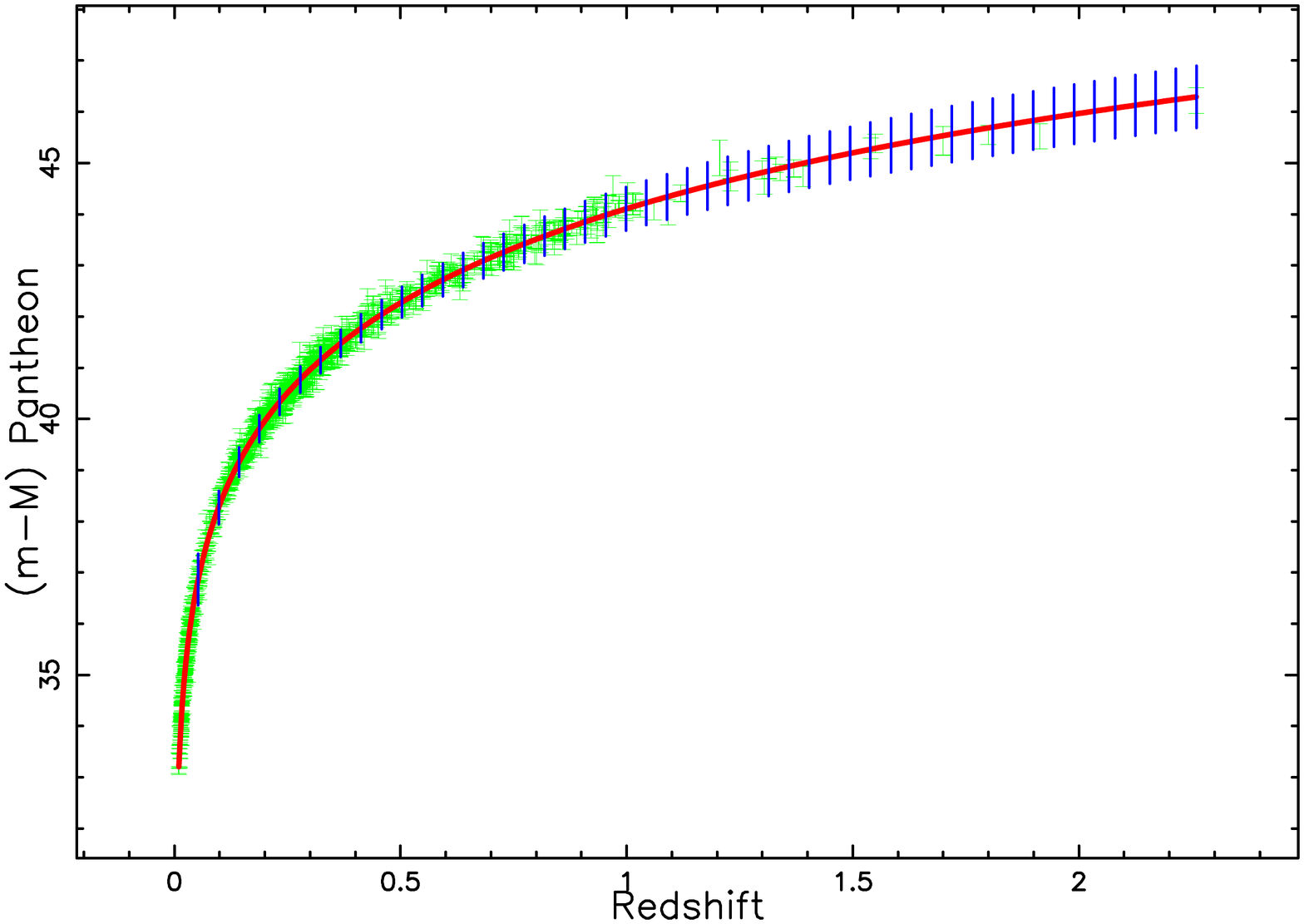}
\caption{
Hubble diagram for the  Pantheon  sample, green points
with error bar.
The solid red line represents the best fit
for the distance modulus  in 
flat   cosmology  
as represented by Eq.~(\ref{distmod_flatnew}).
The theoretical uncertainties are  represented through 
blue vertical lines.
}
\label{distmrev_pantheon}
\end{figure}

The fourth test is performed on the Pantheon 
sample of 1048 SN~Ia \cite{Jones2018,Scolnic2018}
with  
calibrated
data available at
\url{https://archive.stsci.edu/prepds/ps1cosmo/jones_datatable.html},
see Table \ref{chi2_pantheon}
and Figure \ref{distmrev_pantheon} for the 
best fit in the flat  cosmology.
\begin{table}[ht!]
\caption
{
Numerical values of  
$\chi^2$, 
$\chi_{red}^2$, 
$Q$
and the AIC of 
the Hubble diagram for 
the Pantheon sample,
$k$ stands for the number of parameters,
$H_0$ is expressed in 
$\h0units$; 1048 SN~Ia.
}
\label{chi2_pantheon}
\begin{center}
\resizebox{12cm}{!}
{
\begin{tabular}{|c|c|c|c|c|c|c|c|}
\hline
cosmology & Eq. &  $k$         &   parameters    & $\chi2$& $\chi_{red}^2$ &
$Q$  & AIC \\
\hline
$\Lambda$CDM & (\ref{distmod_lcdm}) &  3
& $H_0=  ( 68.209\pm 0.2)$ ; $\om=( 0.278 \pm 0.02) $; 
$\ola= (0.651 \pm 0.02)$
&  1054.71 & 1.01  & 0.41 &    1060.71 \\
\hline
wCDM  
& (\ref{distmod_dark}) 
& 3 
& $H_0=(69.8 \pm  0.27)$; 
$\om =(0.3 \pm  0.016)$ 
; $w=  ( -0.989 \pm   0.03)$ 
&  1053.67 &  1  &  0.419 &  1059.67
\\
\hline
Cardassian 
& (\ref{distmod_cardassian})  
& 3
& $H_0=( 70.01    \pm 0.31   )$; 
$\om  =(  0.329   \pm 0.014  )$ 
; $n=  (  -0.091  \pm  0.005 )$ 
&  1054.49 & 1 &  0.412 
&  1060.49
\\
\hline
flat &(\ref{distmod_flatnew}) &  2
& $H_0=69.94 \pm 0.171 $; $\om=  0.296 \pm  0.002$
&  1053.53  & 1 &  0.429  & 1057.53 \\
\hline
$\phi$CDM   &(\ref{distmod_num_phi}) &  4
& $H_0=69.7\pm 0.1$; $\zerom= 0.28\pm0.02$; $\alpha=0.12 \pm 0.2$; 
$\beta=0.05 \pm 0.02$
& 1053.84 &  1  & 0.4 & 1061.84 \\
\hline
Einstein--De Sitter  & (\ref{distmod_eds}) & 1  &
$H_0=  62.71 \pm    0.2  $
&   2387.62  & 2.28 & 0   &  2389.62\\
\hline 
EdesNa  & (\ref{distmod_edsna}) & 1  &
$H_0=69.1 \pm  0.13 $
&  1059.84 &  1.01 & 0.384  &    1061.8 \\
\hline 
simple~GR  & (\ref{distmod_simplegr}) & 2  &$H_0=73.79\pm  0.015$,
$q_0$=-0.063
&  1476.59 &   1.411 &  $ 2.67\,10^{-17}$ &   1480.59\\
\hline
flat~expanding~model & (\ref{distmod_heymann}) & 1  &
$H_0= 66.67 \pm 0.12 $
&  1219  & 1.16 & $1.6\,10^{-4}$  & 1221 \\
\hline
Milne  &(\ref{distmod_milne}) & 1  &
$H_0=67.37 \pm  0.12 $
&  1132.6 & 1.08 &   0.033 & 1134.6 \\
\hline
plasma    & (\ref{distmod_plasma})& 1 & $H_0= 74.7 \pm 0.14$
&  2017.3 & 1.92   & 0  & 2019.3  \\
\hline
MTL & (\ref{distmod_tired}) & 2 & $\beta$=2.31 , $H_0= 68.95 \pm  0.222$ &
 1069.7 & 1.022 & 0.298 & 1073.7\\
\hline
\end{tabular}
}
\end{center}
\end{table}

\begin{figure}
\begin{center}
\includegraphics[width=6cm]{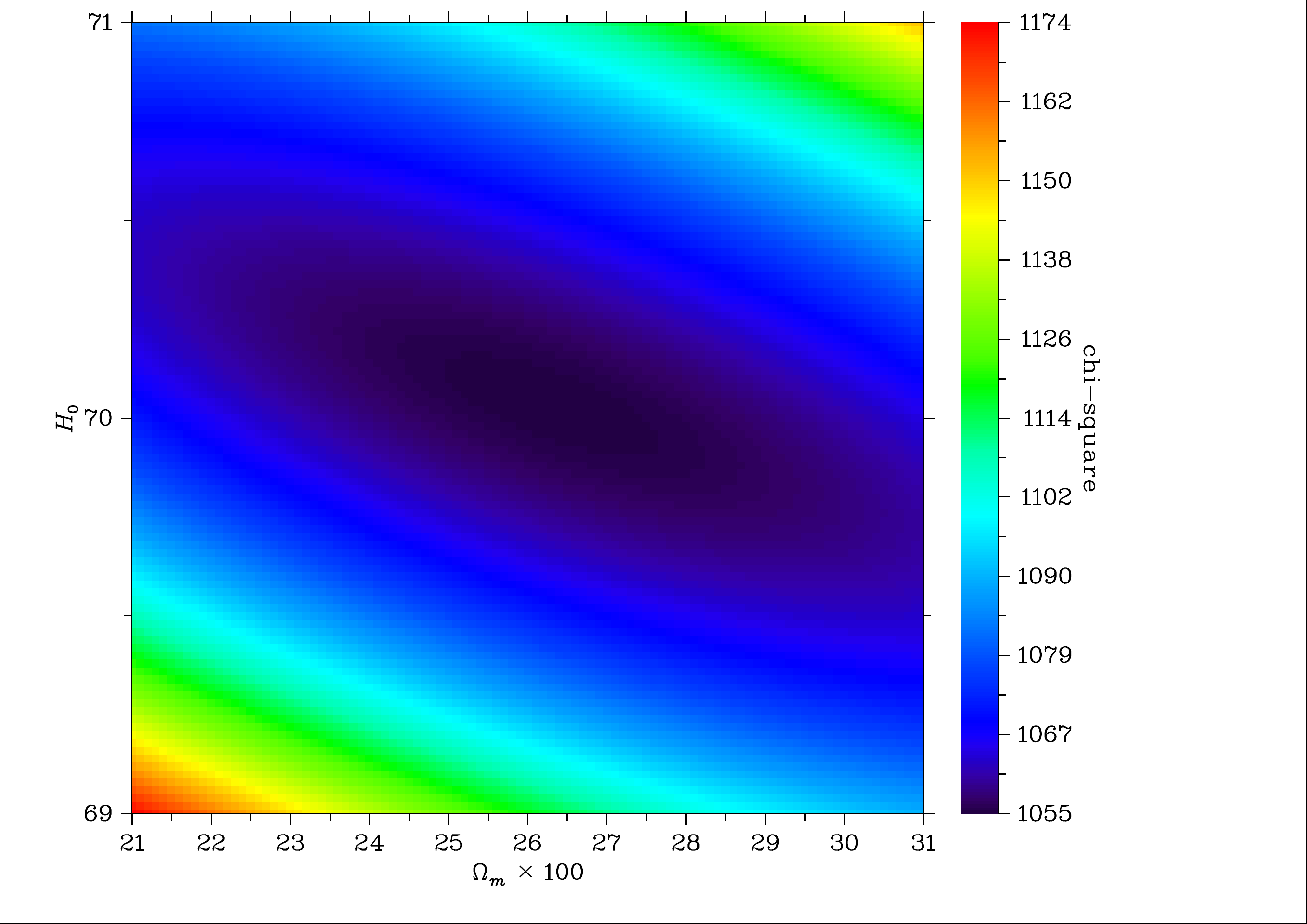}
\end{center}
\caption
{
Color contour plot for $\chi2$  of  the Hubble diagram 
for the Pantheon sample 
in  $\Lambda$CDM cosmology
when $H_0$ and $\om$ are variables and
$\ola=0.626$.}
\label{chi2_dislin}
\end{figure}

In order to see how $\chi^2$ varies around the minimum
for the Pantheon sample
in the case of the $\Lambda$CDM  cosmology,
Figure \ref{chi2_dislin}
presents a 2D colour map for the
values of $\chi^2$ for the Pantheon sample  
when $H_0$ and $\om$ are allowed
to vary around the numerical values which  fix the minimum.

\begin{figure}
\includegraphics[width=6cm]{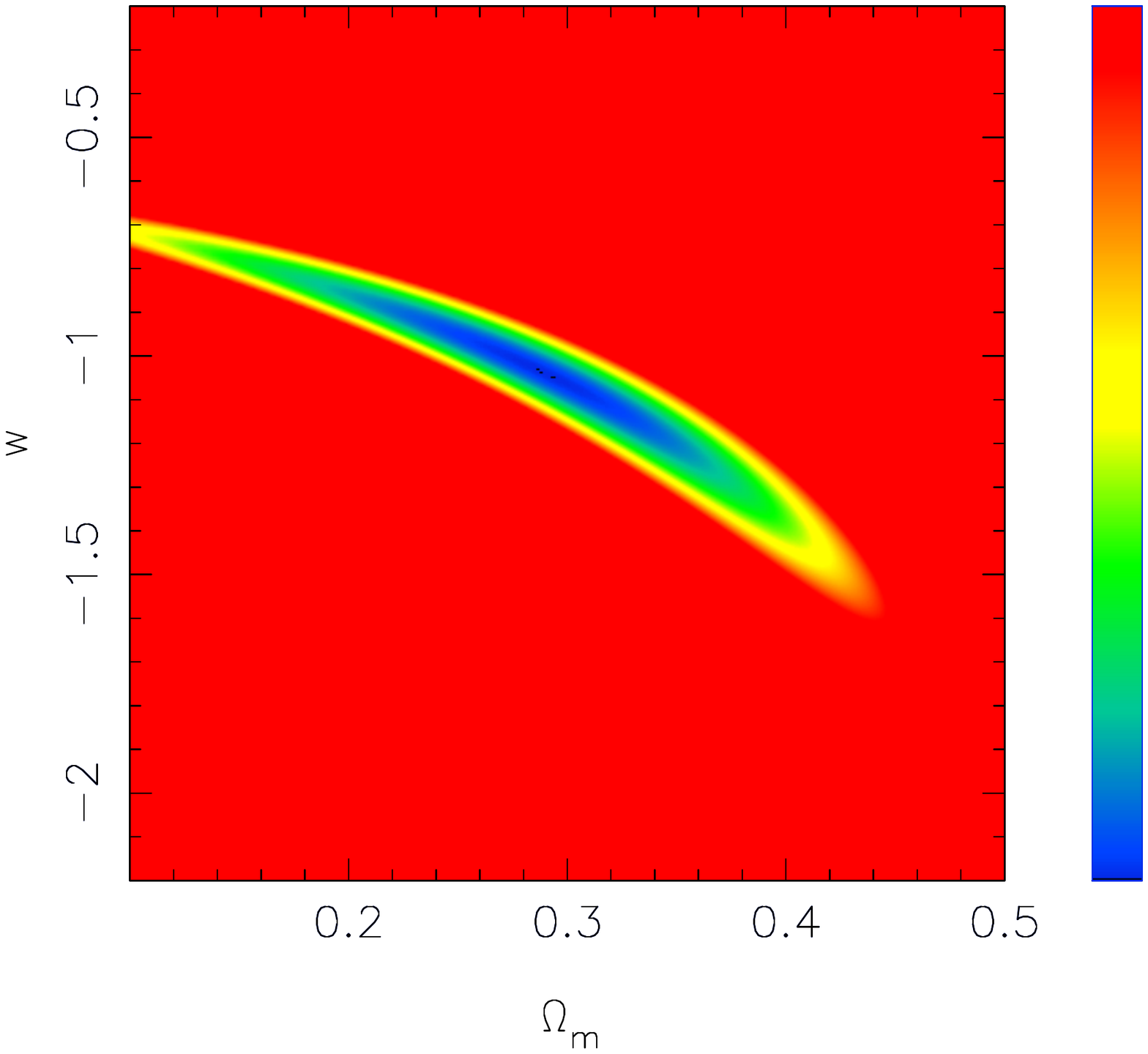}
\caption{
Map of $\chi^2$ 
for the wCDM  cosmology
when  $H_0=69.8$. 
}
\label{dark_color}
\end{figure}

Figure \ref{dark_color} presents 
the map for $\chi^2$ for wCDM for the Pantheon sample 
when  $H_0$ is fixed and  $\om$ and $w$ are allowed to vary.

\subsection{Angular-diameter distance}

In 
the relativistic models the angular 
diameter distance, $\da$ \cite{Etherington1933},
is 
\begin{equation}
\da = \frac{\dl}{(1+z)^2}
\quad .
\end{equation}
We now introduce the minimax  approximation.
Let $f(x)$ be a real function defined
in the interval
$[a, b]$.  The best  rational approximation of degree $(k, l)$
evaluates the coefficients of the ratio of two polynomials
of degree $k$ and $l$, respectively, which minimizes
the maximum difference of
\begin{equation}
max \bigl |f(x) -\frac{p_{0}+p_{1}x+\dots+p_{k}x^{k}}
{q_0+q_{1}x+\dots+q_{\ell}x^{%
\ell}}\bigr |
\quad ,
\end{equation}
on the interval $[a, b]$.
The quality of the fit is  given by the maximum error
over  the considered range.
The coefficients are evaluated through the Remez algorithm,
see \cite{Remez1934,Remez1957}. 
The  minimax  approximation for the angular distance 
in   the interval $0 < z < 8$ 
with data as in Table \ref{chi2_union_grb}
for $\Lambda$CDM cosmology
when   $k=2$ and  $p=2$    
is    
\begin{eqnarray}
D_{A,2,2} =
{\frac {- 0.08126207+ \left(  296.9974312+ 2.715947207\,z \right) z}{
 0.0672056121+ \left(  0.0810298760+ 0.02498056665\,z \right) z}}
\, Mpc
\\
maximum ~error ~=0.6911273~Mpc
\quad ,
\nonumber
\end{eqnarray}
for wCDM cosmology
when  $k=3$ and  $p=2$ is    
\begin{eqnarray}
D_{A,3,2} =
\nonumber  \\
{\frac { 0.034977336+ \left(  287.18685+ \left(  1.1871126+
 0.0002567152\,z \right) z \right) z}{ 0.0665238+ \left( 
 0.09134443+ 0.023282807\,z \right) z}}
\, Mpc
\\
maximum ~error ~=0.07~Mpc
\quad ,
\nonumber
\end{eqnarray}
for Cardassian cosmology
when   $k=2$ and  $p=2$    
is    
\begin{eqnarray}
D_{A,2,2} =
{\frac {- 0.11928613+ \left(  273.3160492+ 2.420885784\,z \right) z}{
 0.0638700712+ \left(  0.0750594027+ 0.02611741351\,z \right) z}}
\, Mpc
\\
maximum ~error ~=0.8346776~Mpc
\quad ,
\nonumber
\end{eqnarray}
for flat cosmology
when   $k=2$ and  $p=2$    
is    
\begin{eqnarray}
D_{A,2,2} =
{\frac {- 0.03653022+ \left(  274.6370918+ 2.192330157\,z \right) z}{
 0.0641307653+ \left(  0.0767316787+ 0.02582682170\,z \right) z}}
 Mpc
\\
maximum ~error ~=0.629004~Mpc
\quad ,
\nonumber
\end{eqnarray}
and for  $\phi$CDM  cosmology
when   $k=2$ and  $p=2$    
is    
\begin{eqnarray}
D_{A,2,2} =
{\frac {- 0.01852238+ \left(  278.5646306+ 2.230340777\,z \right) z}{
 0.0652823706+ \left(  0.0768568011+ 0.02575830541\,z \right) z}}
\\
maximum ~error ~=0.6261293~Mpc
\quad .
\nonumber
\end{eqnarray}
In MTL  there is no difference  between the distance 
$d$ , see equation (\ref{nonlzd}), and the angular distance.
We report  the numerical value of  $d$ 
in   the interval $0 < z < 8$ 
with data as in Table \ref{chi2_union_grb}
\begin{equation}
d= 
4330.383620\,\ln  \left( z+1 \right)  
\, Mpc
\quad . 
\end{equation}

\begin{figure}
\includegraphics[width=6cm]{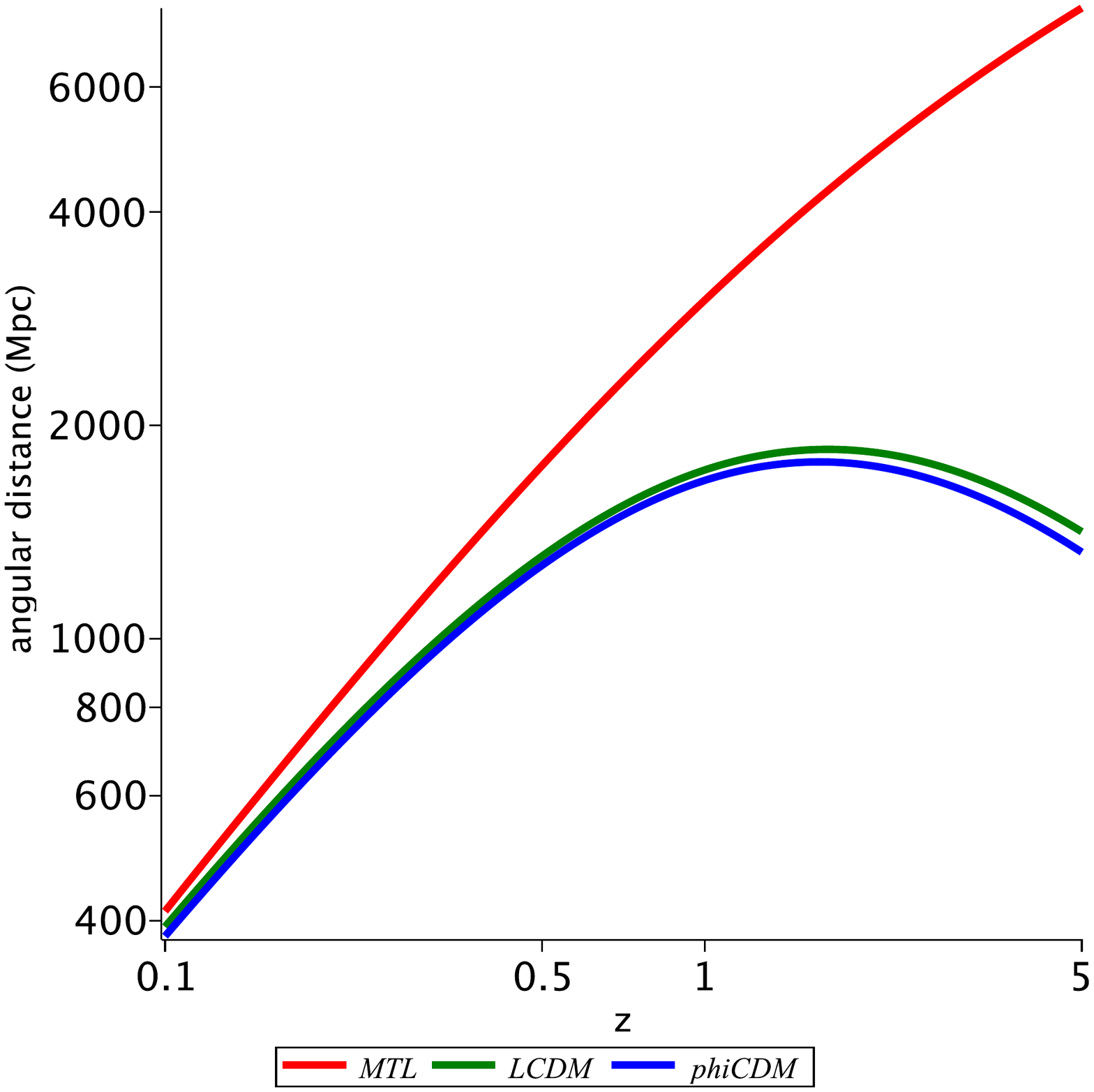}
\caption{
Angular  distance in  
MTL (red) ,  
$\Lambda$CDM (green)
and 
$\phi$CDM (blue) cosmologies 
with data as in Table 
\ref{chi2_union_grb}
}
\label{mtlothers}
\end{figure}

A promising field of investigation in applied cosmology
is the  maximum  of the angular distance as function of the 
redshift \cite{Braatz2010,Kuo2013}, 
$z_{max}$, which is finite in relativistic cosmologies
and  infinite in the Milne, plasma and MTL cosmologies,
see  Figure \ref{mtlothers}. 
The  numerical value  of $z_{max}$ 
is  reported in  Table \ref{tablezmax},
as  a reference  $z_{max}=1.594$ for flat Planck$\Lambda$CDM 
cosmology \cite{Melia2018}.

\begin{table}[ht!]
\caption
{
Numerical values of  $z_{max}$  and
radius of Einstein ring  in kpc
when $R_{\mathrm {ave}} = 1.54\,arcsec$
}
\label{tablezmax}
\begin{center}
\begin{tabular}{|c|c|c|}
\hline
cosmology &  $z_{max}$ & radius (kpc)  \\
\hline
$\Lambda$CDM  & 1.691  &  13.333   \\ 
\hline
wCDM          & 1.716   &  11.797  \\
\hline
Cardassian    & 1.607   & 11.938 \\ 
\hline        
flat          & 1.615  & 11.907  \\ 
\hline
$\phi$CDM     & 1.632  & 12.05   \\
\hline 
MTL           & $\infty$  & 45.15 \\
\hline
\end{tabular}
\end{center}
\end{table}
Another example is given  by 
the ring associated with the galaxy SDP.81,
see \cite{Eales2010}, 
 which is   generally explained by the gravitational lens.
In this framework we have a foreground galaxy at 
$z=0.2999$ and a background galaxy at $z=0.3042$.
This ring  has  been studied with the 
Atacama Large Millimeter/sub-millimeter Array (ALMA)
by 
\cite{Tamura2015,ALMA2015,Rybak2015,Hatsukade2015,Wong2015,Hezaveh2016}.
The system SDP.81 
as been analysed 
by  ALMA  and presents 14  molecular clumps along the two main
lensed arcs: the  averaged  
radius in $arcsec$
is   $R_{\mathrm {ave}} = 1.54\,arcsec$ \cite{Zaninetti2017c}.

\section{Conclusions}

{\bf Cosmological models}
We list according to increasing order of the values of the  merit function, $\chi^2$,
the first, second, third and fourth    
cosmological models, see  Table \ref{bestfit}.

\begin{table}[ht!]
\caption
{
The first, second, third and fourth best fitting models for the 
four compilations. 
}
\label{bestfit}
\begin{center}
\resizebox{12cm}{!}
{
\begin{tabular}{|c|c|c|c|c|}
\hline
Compilation & first~model & second~model & third model &fourth model\\
\hline
Union~2.1       & wCDM Hypergeometric    & Cardassian & 
$\phi$CDM       & flat
\\
JLA              & wCDM Hypergeometric   & MTL  & $\phi$CDM  & $\Lambda$CDM
\\
Union~2.1+GRBs  & $\Lambda$CDM          &$\phi$CDM  & Cardassian &  flat
\\
Pantheon         & wCDM Hypergeometric   & Cardassian  
& flat & $\phi$CDM 
\\
\hline
\end{tabular}
}
\end{center}
\end{table}

The  Einstein--De Sitter, simple GR and plasma models produce
the highest values in the $\chi^2$
and are here considered only for historical reasons.

\begin{figure}
\begin{center}
\includegraphics[width=8cm]{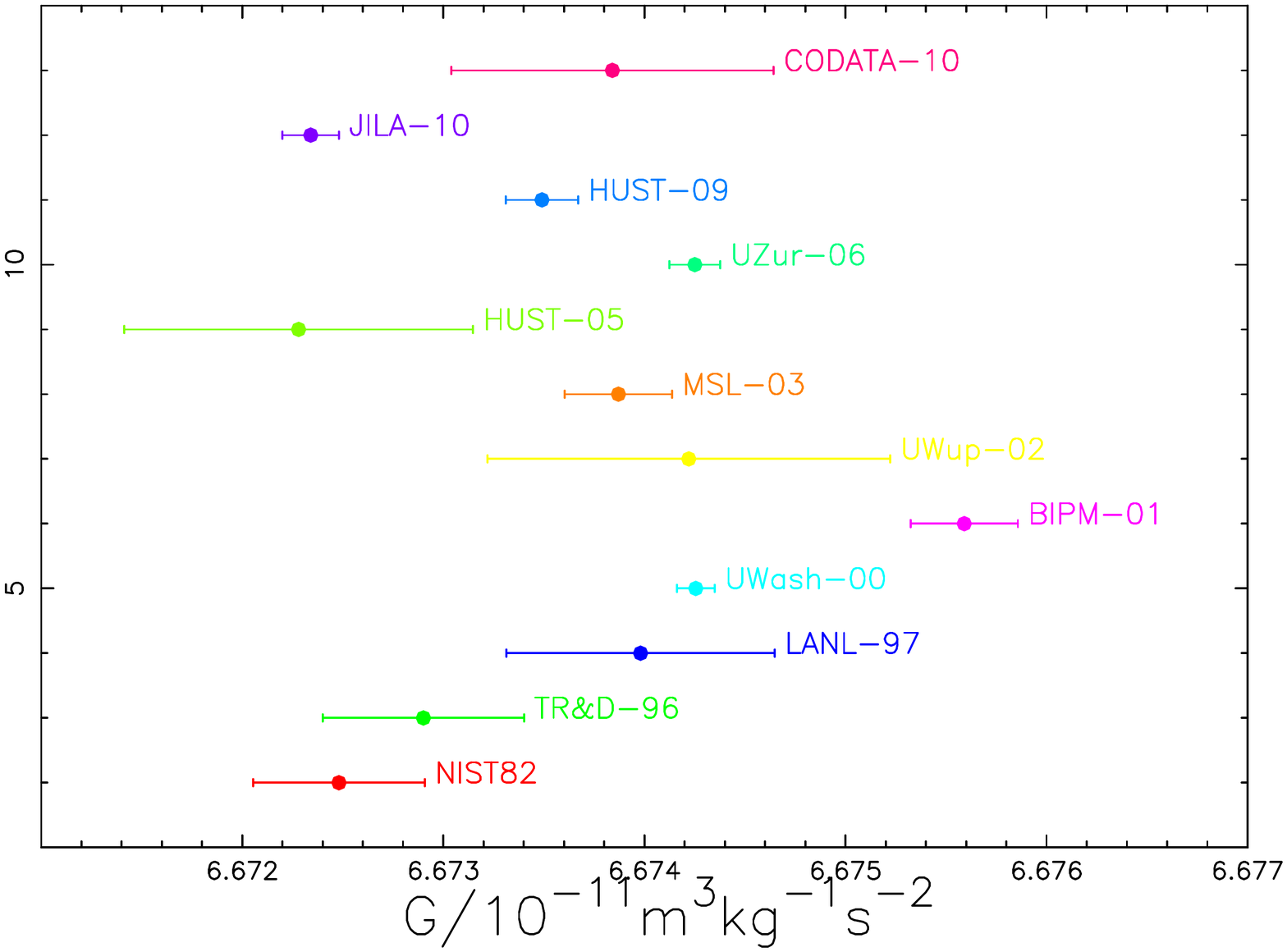}
\end{center}
\caption
{
Values of the Newtonian constant of gravitation
$G$ as  given by Table XXIV   in  
\cite{CODATA2012}.
}
\label{table_g_great}
\end{figure}

\begin{figure}
\begin{center}
\includegraphics[width=8cm]{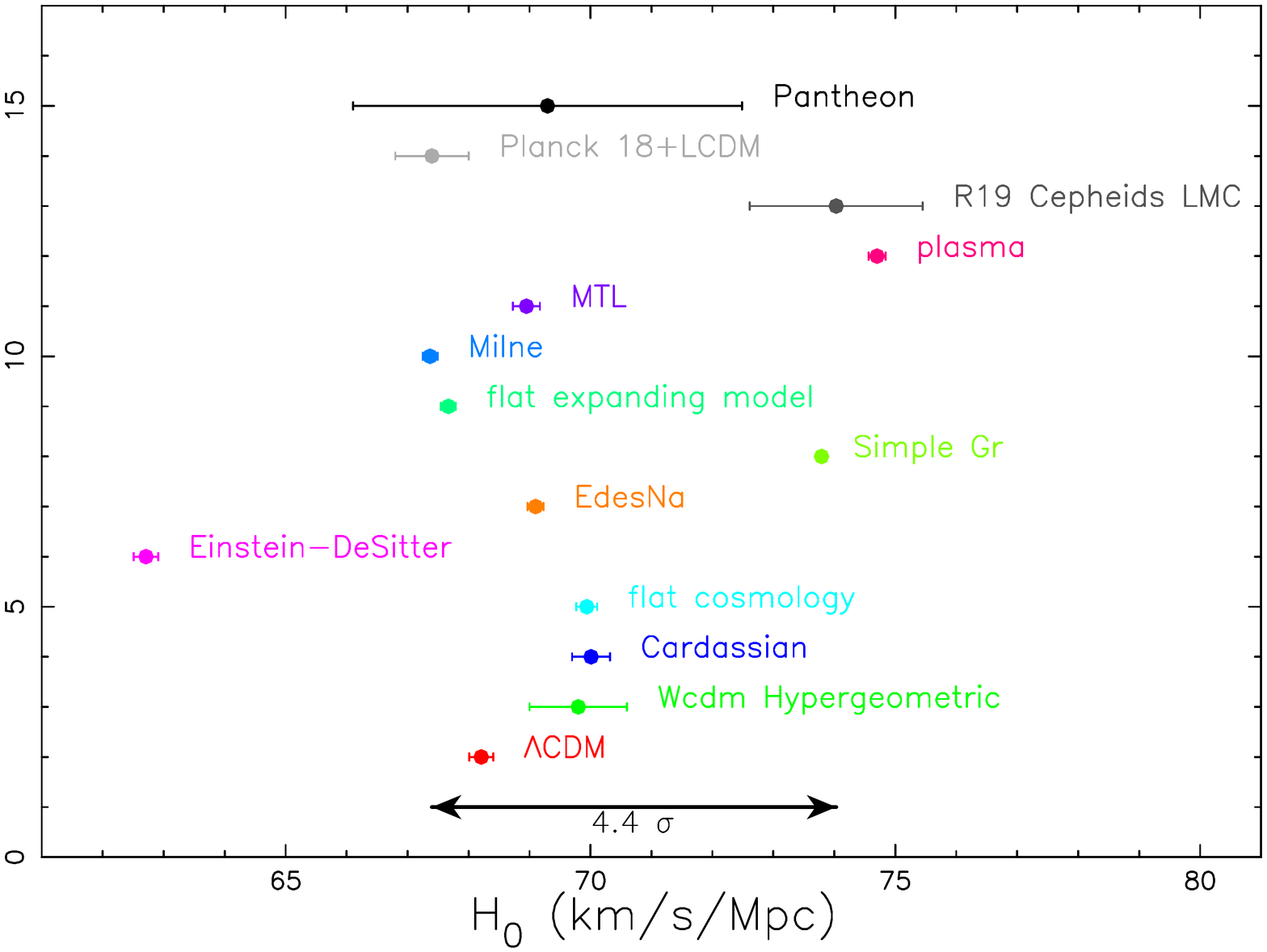}
\end{center}
\caption
{
The present tension on $H_0$ (black line with two arrows) 
and our results 
in the case of the Pantheon sample
with the connected averaged value which is marked as `Pantheon'; 
parameters as in  Table \ref{chi2_pantheon}.
}
\label{distmrev_table}
\end{figure}

{\bf Physics versus Astronomy}
The
value of the Newtonian gravitational constant, denoted by $G$,
is derived applying the weighted mean, but the uncertainties 
were multiplied
by a factor of 14, 
of 11 values  available in Table XXIV in
\cite{CODATA2012},
see  Figure  \ref{table_g_great}.
By analogy, we  average the  values  of $H_0$ for the Pantheon 
sample  and 
we report as  error for $H_0$ the standard deviation
\begin{equation}
\overline{H_0} = (69.29 \pm 3.18 ) 
\h0units 
\quad Pantheon~ sample \quad,
\end{equation}
see Figure \ref{distmrev_table}. 

\appendix
\setcounter{equation}{0}
\renewcommand{\theequation}{\thesection.\arabic{equation}}

\section{The Pad\'e approximant}
\label{appendixa}
Given a function $f(z)$, the Pad\'e  approximant,
after \cite{Pade1892},
is
\begin{equation}
f(z)=\frac{a_{0}+a_{1}z+\dots+a_{p}z^{p}}{b_{0}+b_{1}%
z+\dots+b_{q}z^{q}}
\quad ,
\end{equation}
where the notation is the same as in \cite{NIST2010}.

The coefficients $a_i$ and $b_i$
are found through Wynn's cross rule,
see \cite{Baker1975,Baker1996}
and our choice is $p=2$ and $q=2$.
The choice of  $p$ and $q$ is a compromise between
precision (associated with high values for  $p$ and $q$) and
the simplicity of the expressions to manage
(associated with low values for  $p$ and $q$).
The argument of the integral to be done  
is the inverse of $E(z)$, see  Eq.~(\ref{eq:ez}), 
\begin{equation}
\label{argumentint}
\frac{1}{E(z)} =\frac{1}{ \sqrt{\om\,(1+z)^3+\ok\,(1+z)^2+\ola}}
\quad ,
\end{equation}
and the Pad\'e  approximant  is
\begin{equation}
\label{argumentintpade}
\frac{1}{E(z)} =
\frac{a_0 + a_1 z + a_2 z^2}{b_0 + b_1 z + b_2 z^2}
\quad ,
\end{equation}
where
\begin{eqnarray}
a_0 =
16\, \bigl( 32\,{{\it \ok}}^{3}{\it \ola}+16\,{{\it \ok}}^{2}{{\it
\ola}}^{2}+160\,{{\it \ok}}^{2}{\it \ola}\,{\it \om}+24\,{{\it \ok}}^{
2}{{\it \om}}^{2}
\nonumber \\
+64\,{\it \ok}\,{{\it \ola}}^{2}{\it \om}
+320\,{\it
\ok}\,{\it \ola}\,{{\it \om}}^{2}+40\,{\it \ok}\,{{\it \om}}^{3}+96\,{
{\it \ola}}^{2}{{\it \om}}^{2}+
\nonumber \\
192\,{\it \ola}\,{{\it \om}}^{3}
+15\,{{
\it \om}}^{4}\bigr )  \bigl( {\it \om}+{\it \ok}+{\it \ola} \bigr) ^
{4}
\end{eqnarray}
\begin{eqnarray}
a_1 =
4\, \bigr ( 128\,{{\it \ok}}^{4}{\it \ola}+32\,{{\it \ok}}^{3}{{\it
\ola}}^{2}+704\,{{\it \ok}}^{3}{\it \ola}\,{\it \om}
-16\,{{\it \ok}}^{
2}{{\it \ola}}^{2}{\it \om}
\nonumber \\
+1456\,{{\it \ok}}^{2}{\it \ola}\,{{\it \om
}}^{2}+32\,{{\it \ok}}^{2}{{\it \om}}^{3}-64\,{\it \ok}\,{{\it \ola}}^
{3}{\it \om}-384\,{\it \ok}\,{{\it \ola}}^{2}{{\it \om}}^{2}
\nonumber \\
+1512\,{
\it \ok}\,{\it \ola}\,{{\it \om}}^{3}
+50\,{\it \ok}\,{{\it \om}}^{4}-
192\,{{\it \ola}}^{3}{{\it \om}}^{2}-288\,{{\it \ola}}^{2}{{\it \om}}^
{3}
\nonumber  \\
+648\,{\it \ola}\,{{\it \om}}^{4}
+15\,{{\it \om}}^{5} \bigr )
 \bigl( {\it \om}+{\it \ok}+{\it \ola} \bigr) ^{3}
\end{eqnarray}
\begin{eqnarray}
a_2 =
- \bigl( 256\,{{\it \ok}}^{4}{\it \ola}\,{\it \om}-64\,{{\it \ok}}^{3}
{{\it \ola}}^{3}+320\,{{\it \ok}}^{3}{{\it \ola}}^{2}{\it \om}
\nonumber \\
+960\,{{
\it \ok}}^{3}{\it \ola}\,{{\it \om}}^{2}
-320\,{{\it \ok}}^{2}{{\it
\ola}}^{3}{\it \om}+240\,{{\it \ok}}^{2}{{\it \ola}}^{2}{{\it \om}}^{2
}
\nonumber \\
+1440\,{{\it \ok}}^{2}{\it \ola}\,{{\it \om}}^{3}
+16\,{{\it \ok}}^{2}
{{\it \om}}^{4}
-1600\,{\it \ok}\,{{\it \ola}}^{3}{{\it \om}}^{2}-480\,
{\it \ok}\,{{\it \ola}}^{2}{{\it \om}}^{3}
\nonumber \\
+1140\,{\it \ok}\,{\it \ola}
\,{{\it \om}}^{4}
+20\,{\it \ok}\,{{\it \om}}^{5}
-256\,{{\it \ola}}^{4}
{{\it \om}}^{2}
-1600\,{{\it \ola}}^{3}{{\it \om}}^{3}
\nonumber \\
-240\,{{\it \ola}
}^{2}{{\it \om}}^{4}
+380\,{\it \ola}\,{{\it \om}}^{5}
+5\,{{\it \om}}^{
6} \bigr)  \bigl( {\it \om}+{\it \ok}+{\it \ola} \bigr) ^{2}
\end{eqnarray}
\begin{eqnarray}
b_0 = 16\, \bigl( {\it \om}+{\it \ok}+{\it \ola} \bigr) ^{9/2} \bigl( 32\,{
{\it \ok}}^{3}{\it \ola}+16\,{{\it \ok}}^{2}{{\it \ola}}^{2}+160\,{{
\it \ok}}^{2}{\it \ola}\,{\it \om}
\nonumber \\
+24\,{{\it \ok}}^{2}{{\it \om}}^{2}+
64\,{\it \ok}\,{{\it \ola}}^{2}{\it \om}+320\,{\it \ok}\,{\it \ola}\,{
{\it \om}}^{2}+40\,{\it \ok}\,{{\it \om}}^{3}
\nonumber \\
+96\,{{\it \ola}}^{2}{{
\it \om}}^{2}
+192\,{\it \ola}\,{{\it \om}}^{3}+15\,{{\it \om}}^{4}
 \bigr)
\end{eqnarray}
\begin{eqnarray}
b_1 =
4\, \bigl( {\it \om}+{\it \ok}+{\it \ola} \bigr) ^{7/2} \bigl( 256\,{
{\it \ok}}^{4}{\it \ola}+96\,{{\it \ok}}^{3}{{\it \ola}}^{2}+1536\,{{
\it \ok}}^{3}{\it \ola}\,{\it \om}
\nonumber  \\
+96\,{{\it \ok}}^{3}{{\it \om}}^{2}+
336\,{{\it \ok}}^{2}{{\it \ola}}^{2}{\it \om}+3696\,{{\it \ok}}^{2}{
\it \ola}\,{{\it \om}}^{2}
\nonumber \\
+336\,{{\it \ok}}^{2}{{\it \om}}^{3}-64\,{
\it \ok}\,{{\it \ola}}^{3}{\it \om}+384\,{\it \ok}\,{{\it \ola}}^{2}{{
\it \om}}^{2}+4200\,{\it \ok}\,{\it \ola}\,{{\it \om}}^{3}
\nonumber \\
+350\,{\it
\ok}\,{{\it \om}}^{4}
-192\,{{\it \ola}}^{3}{{\it \om}}^{2}
+288\,{{\it
\ola}}^{2}{{\it \om}}^{3}+1800\,{\it \ola}\,{{\it \om}}^{4}+105\,{{
\it \om}}^{5} \bigr)
\end{eqnarray}
\begin{eqnarray}
b_2 =
 \bigl( {\it \om}+{\it \ok}+{\it \ola} \bigr) ^{5/2} \bigl( 512\,{{
\it \ok}}^{5}{\it \ola}+384\,{{\it \ok}}^{4}{{\it \ola}}^{2}+3584\,{{
\it \ok}}^{4}{\it \ola}\,{\it \om}
\nonumber \\
+192\,{{\it \ok}}^{3}{{\it \ola}}^{3
}+1984\,{{\it \ok}}^{3}{{\it \ola}}^{2}{\it \om}+10752\,{{\it \ok}}^{3
}{\it \ola}\,{{\it \om}}^{2}+320\,{{\it \ok}}^{3}{{\it \om}}^{3}
\nonumber \\
+960\,
{{\it \ok}}^{2}{{\it \ola}}^{3}{\it \om}
+5136\,{{\it \ok}}^{2}{{\it
\ola}}^{2}{{\it \om}}^{2}+17760\,{{\it \ok}}^{2}{\it \ola}\,{{\it \om}
}^{3}
\nonumber \\
+840\,{{\it \ok}}^{2}{{\it \om}}^{4}
+2752\,{\it \ok}\,{{\it \ola}
}^{3}{{\it \om}}^{2}
+7392\,{\it \ok}\,{{\it \ola}}^{2}{{\it \om}}^{3}+
\nonumber \\
15060\,{\it \ok}\,{\it \ola}\,{{\it \om}}^{4}
\nonumber \\
+700\,{\it \ok}\,{{\it
\om}}^{5}
+256\,{{\it \ola}}^{4}{{\it \om}}^{2}+2752\,{{\it \ola}}^{3}{
{\it \om}}^{3}
\nonumber \\
+3696\,{{\it \ola}}^{2}{{\it \om}}^{4}
+5020\,{\it \ola}
\,{{\it \om}}^{5}+175\,{{\it \om}}^{6} \bigr).
\end{eqnarray}

The  indefinite integral  of (\ref{argumentintpade}), $F_{2,2}$, is 
\begin{eqnarray}
F_{2,2}(z;a_0,a_1,a_2,b_0,b_1,b_2) =
{\frac {a_{{2}}z}{b_{{2}}}} \nonumber \\
+\frac{1}{2}\,{\frac {\ln  \left( {z}^{2}b_{{2}}+zb
_{{1}}+b_{{0}} \right) a_{{1}}}{b_{{2}}}}-\frac{1}{2}\,{\frac {\ln  \left( {z}
^{2}b_{{2}}+zb_{{1}}+b_{{0}} \right) a_{{2}}b_{{1}}}{{b_{{2}}}^{2}}}
\nonumber \\
+2
\,{\frac {a_{{0}}}{\sqrt {4\,b_{{0}}b_{{2}}-{b_{{1}}}^{2}}}\arctan
 \left( {\frac {2\,zb_{{2}}+b_{{1}}}{\sqrt {4\,b_{{0}}b_{{2}}-{b_{{1}}
}^{2}}}} \right) }
\nonumber \\
-2\,{\frac {a_{{2}}b_{{0}}}{b_{{2}}\sqrt {4\,b_{{0}}
b_{{2}}
-{b_{{1}}}^{2}}}\arctan \left( {\frac {2\,zb_{{2}}+b_{{1}}}{
\sqrt {4\,b_{{0}}b_{{2}}-{b_{{1}}}^{2}}}} \right) }
\nonumber \\
-{\frac {b_{{1}}a_{
{1}}}{b_{{2}}\sqrt {4\,b_{{0}}b_{{2}}-{b_{{1}}}^{2}}}\arctan \left( {
\frac {2\,zb_{{2}}+b_{{1}}}{\sqrt {4\,b_{{0}}b_{{2}}-{b_{{1}}}^{2}}}}
 \right) }
\nonumber \\
+{\frac {{b_{{1}}}^{2}a_{{2}}}{{b_{{2}}}^{2}\sqrt {4\,b_{{0}
}b_{{2}}-{b_{{1}}}^{2}}}\arctan \left( {\frac {2\,zb_{{2}}+b_{{1}}}{
\sqrt {4\,b_{{0}}b_{{2}}-{b_{{1}}}^{2}}}} \right) }
\quad  .
\label{integral22}
\end{eqnarray}

\section*{Acknowledgments}

The author is grateful 
to  David Jones for information useful 
for downloading the data of the Pantheon sample.
 

\begin{thebibliography}{10}
\expandafter\ifx\csname url\endcsname\relax
  \def\url#1{{\tt #1}}\fi
\expandafter\ifx\csname urlprefix\endcsname\relax\def\urlprefix{URL }\fi
\providecommand{\eprint}[2][]{\url{#2}}

\bibitem{Planck2018}
{Planck Collaboration}, {Aghanim} N, {Akrami} Y and et~al 2018 {Planck 2018
  results. VI. Cosmological parameters} {\em ArXiv e-prints\/}
  (\textit{Preprint} \eprint{1807.06209})

\bibitem{Riess2019}
{Riess} A~G, {Casertano} S, {Yuan} W, {Macri} L~M and {Scolnic} D 2019 {Large
  Magellanic Cloud Cepheid Standards Provide a 1\% Foundation for the
  Determination of the Hubble Constant and Stronger Evidence for Physics beyond
  {\ensuremath{\Lambda}}CDM} {\em \apj\/} {\bf 876}(1) 85 (\textit{Preprint}
  \eprint{1903.07603})

\bibitem{DiValentino2020}
{Di Valentino} E, {Anchordoqui} L~A, {Akarsu} O and et~al 2020 {Cosmology
  Intertwined II: The Hubble Constant Tension} {\em arXiv e-prints\/}
  arXiv:2008.11284 (\textit{Preprint} \eprint{2008.11284})

\bibitem{Riess1998}
{Riess} A~G, {Filippenko} A~V, {Challis} P and {Clocchiatti} A 1998
  {Observational Evidence from Supernovae for an Accelerating Universe and a
  Cosmological Constant} {\em \aj\/} {\bf 116}, 1009 (\textit{Preprint}
  \eprint{astro-ph/9805201})

\bibitem{Suzuki2012}
{Suzuki} N, {Rubin} D, {Lidman} C, {Aldering} G, {Amanullah} R, {Barbary} K and
  {Barrientos} L~F 2012 {The Hubble Space Telescope Cluster Supernova Survey.
  V. Improving the Dark-energy Constraints above z greater than 1 and Building
  an Early-type-hosted Supernova Sample} {\em \apj\/} {\bf 746} 85

\bibitem{Betoule2014}
{Betoule} M, {Kessler} R, {Guy} J and {Mosher} J 2014 {Improved cosmological
  constraints from a joint analysis of the SDSS-II and SNLS supernova samples}
  {\em \aap\/} {\bf 568} A22

\bibitem{Jones2018}
{Jones} D~O, {Scolnic} D~M, {Riess} A~G and et~al 2018 {Measuring Dark Energy
  Properties with Photometrically Classified Pan-STARRS Supernovae. II.
  Cosmological Parameters} {\em \apj\/} {\bf 857}(1) 51 (\textit{Preprint}
  \eprint{1710.00846})

\bibitem{Scolnic2018}
{Scolnic} D~M, {Jones} D~O, {Rest} A and et~al 2018 {The Complete Light-curve
  Sample of Spectroscopically Confirmed SNe Ia from Pan-STARRS1 and
  Cosmological Constraints from the Combined Pantheon Sample} {\em \apj\/} {\bf
  859}(2) 101 (\textit{Preprint} \eprint{1710.00845})

\bibitem{Oliveira2016}
{Oliveira} F~J 2016 {Cosmic Time Transformations in Cosmological Relativity}
  {\em Journal of High Energy Physics, Gravitation and Cosmology\/} {\bf 2},
  253

\bibitem{Gupta2018}
{Gupta} R 2018 {SNe Ia Redshift in a Nonadiabatic Universe} {\em Universe\/}
  {\bf 4}, 104 (\textit{Preprint} \eprint{1810.12090})

\bibitem{Amarzguioui2006}
{Amarzguioui} M, {Elgar{\o}y} {\O}, {Mota} D~F and {Multam{\"a}ki} T 2006
  {Cosmological constraints on f(R) gravity theories within the Palatini
  approach} {\em \aap\/} {\bf 454}(3), 707 (\textit{Preprint}
  \eprint{astro-ph/0510519})

\bibitem{Odintsov2019}
{Odintsov} S~D, {G{\'o}mez} D~S~C and {Sharov} G~S 2019 {Testing logarithmic
  corrections to R$^{2}$-exponential gravity by observational data} {\em
  \prd\/} {\bf 99}(2) 024003

\bibitem{Corda2009}
{Corda} C 2009 {Interferometric Detection of Gravitational Waves: . the
  Definitive Test for General Relativity} {\em International Journal of Modern
  Physics D\/} {\bf 18}(14), 2275 (\textit{Preprint} \eprint{0905.2502})

\bibitem{Lin2019}
{Lin} H~N, {Li} X and {Tang} L 2019 {Non-parametric reconstruction of dark
  energy and cosmic expansion from the Pantheon compilation of type Ia
  supernovae} {\em Chinese Physics C\/} {\bf 43}(7) 075101 (\textit{Preprint}
  \eprint{1905.11593})

\bibitem{Camlibel2020}
Caml{\i}bel A~K, Semiz I and Feyizo{\u{g}}lu M 2020 Pantheon update on a
  model-independent analysis of cosmological supernova data {\em arXiv preprint
  arXiv:2001.04408\/}

\bibitem{Hogg1999}
{Hogg} D~W 1999 {Distance measures in cosmology} {\em ArXiv Astrophysics
  e-prints\/} (\textit{Preprint} \eprint{astro-ph/9905116})

\bibitem{Peebles1993}
{Peebles} P~J~E 1993 {\em {Principles of Physical Cosmology}\/} ({Princeton,
  N.J.}: Princeton University Press)

\bibitem{Zaninetti2016a}
{Zaninetti} L 2016 Pade approximant and minimax rational approximation in
  standard cosmology {\em Galaxies\/} {\bf 4}(1), 4 ISSN 2075-4434
  \urlprefix\url{http://www.mdpi.com/2075-4434/4/1/4}

\bibitem{Turner1997}
{Turner} M~S and {White} M 1997 {CDM models with a smooth component} {\em
  \prd\/} {\bf 56}(8), R4439 (\textit{Preprint} \eprint{astro-ph/9701138})

\bibitem{Tripathi2017}
{Tripathi} A, {Sangwan} A and {Jassal} H~K 2017 {Dark energy equation of state
  parameter and its evolution at low redshift} {\em \jcap\/} {\bf 6} 012
  (\textit{Preprint} \eprint{1611.01899})

\bibitem{Wei2015}
{Wei} J~J, {Ma} Q~B and {Wu} X~F 2015 {Utilizing the Updated Gamma-Ray Bursts
  and Type Ia Supernovae to Constrain the Cardassian Expansion Model and Dark
  Energy} {\em Advances in Astronomy\/} {\bf 2015} 576093 (\textit{Preprint}
  \eprint{1504.02308})

\bibitem{Abramowitz1965}
{Abramowitz} M and {Stegun} I~A 1965 {\em {Handbook of Mathematical Functions
  with Formulas, Graphs, and Mathematical Tables}\/} (New York: Dover)

\bibitem{Seggern1992}
{von Seggern} D 1992 {\em CRC Standard Curves and Surfaces\/} (New York: CRC)

\bibitem{Thompson1997}
{Thompson} W~J 1997 {\em Atlas for computing mathematical functions\/} (New
  York: Wiley-Interscience)

\bibitem{Gradshteyn2007}
{{Gradshteyn}, I~S and {Ryzhik}, I~M and {Jeffrey}, A and {Zwillinger}, D} 2007
  {\em {Table of Integrals, Series, and Products}\/} (New York: Academic Press)

\bibitem{NIST2010}
Olver F~W~J, Lozier D~W, Boisvert R~F and Clark C~W 2010 {\em {NIST {Handbook}
  of {Mathematical} {Functions}}\/} (Cambridge: {Cambridge University Press. })

\bibitem{Zaninetti2019c}
{Zaninetti} L 2019 {The Distance Modulus in Dark Energy and Cardassian
  Cosmologies via the Hypergeometric Function} {\em International Journal of
  Astronomy and Astrophysics\/} {\bf 9}(3), 231 (\textit{Preprint}
  \eprint{1909.02360})

\bibitem{Freese2002}
{Freese} K and {Lewis} M 2002 {Cardassian expansion: a model in which the
  universe is flat, matter dominated, and accelerating} {\em Physics Letters
  B\/} {\bf 540}, 1 (\textit{Preprint} \eprint{astro-ph/0201229})

\bibitem{Freese2003}
{Freese} K 2003 {Generalized cardassian expansion: a model in which the
  universe is flat, matter dominated, and accelerating} {\em Nuclear Physics B
  Proceedings Supplements\/} {\bf 124}, 50 (\textit{Preprint}
  \eprint{hep-ph/0208264})

\bibitem{Baes2017}
{Baes} M, {Camps} P and {Van De Putte} D 2017 {Analytical expressions and
  numerical evaluation of the luminosity distance in a flat cosmology} {\em
  \mnras\/} {\bf 468}, 927 (\textit{Preprint} \eprint{1702.08860})

\bibitem{Zaninetti2019b}
{Zaninetti} L 2019 {A New Analytical Solution for the Distance Modulus in Flat
  Cosmology} {\em International Journal of Astronomy and Astrophysics\/} {\bf
  9}(1), 51 (\textit{Preprint} \eprint{1903.07121})

\bibitem{Starobinsky1980}
{Starobinsky} A~A 1980 {A new type of isotropic cosmological models without
  singularity} {\em Physics Letters B\/} {\bf 91}(1), 99

\bibitem{Guth1981}
{Guth} A~H and {Weinberg} E~J 1981 {Cosmological consequences of a first-order
  phase transition in the SU$_{5}$ grand unified model} {\em \prd\/} {\bf
  23}(4), 876

\bibitem{Ratra1988}
{Ratra} B and {Peebles} P~J~E 1988 {Cosmological consequences of a rolling
  homogeneous scalar field} {\em \prd\/} {\bf 37}(12), 3406

\bibitem{Steinhardt1998}
{Steinhardt} P~J and {Caldwell} R~R 1998 {Introduction to Quintessence} in Y~I
  {Byun} and K~W {Ng}, eds, {\em Cosmic Microwave Background and Large Scale
  Structure of the Universe\/} vol 151 of {\em Astronomical Society of the
  Pacific Conference Series\/} p~13

\bibitem{Avsajanishvili2018}
{Avsajanishvili} O, {Huang} Y, {Samushia} L and {Kahniashvili} T 2018 {The
  observational constraints on the flat {\ensuremath{\varphi}} CDM models} {\em
  European Physical Journal C\/} {\bf 78}(9) 773 (\textit{Preprint}
  \eprint{1711.11465})

\bibitem{Mamon2017}
{Mamon} A~A, {Bamba} K and {Das} S 2017 {Constraints on reconstructed dark
  energy model from SN Ia and BAO/CMB observations} {\em European Physical
  Journal C\/} {\bf 77}(1) 29 (\textit{Preprint} \eprint{1607.06631})

\bibitem{Einstein1932}
{Einstein} A and {de Sitter} W 1932 {On the Relation between the Expansion and
  the Mean Density of the Universe} {\em Proceedings of the National Academy of
  Science\/} {\bf 18}, 213

\bibitem{Krisciunas1993}
{Krisciunas} K 1993 {Look-Back Time the Age of the Universe and the Case for a
  Positive Cosmological Constant} {\em \jrasc\/} {\bf 87}, 223
  (\textit{Preprint} \eprint{astro-ph/9306002})

\bibitem{Ryden2003}
{Ryden} B 2003 {\em {Introduction to Cosmology}\/} (San Francisco, CA, USA:
  Addison Wesley)

\bibitem{Lang2013}
Lang K 2013 {\em Astrophysical Formulae: Space, Time, Matter and Cosmology\/}
  Astronomy and Astrophysics Library (Berlin: Springer) ISBN 9783662216392

\bibitem{Heymann2013}
{Heymann} Y 2013 On the luminosity distance and the hubble constant {\em
  Progress in Physics\/} {\bf 3}, 5

\bibitem{Milne1933}
{Milne} E~A 1933 {World-Structure and the Expansion of the Universe.} {\em
  \zap\/} {\bf 6}, 1

\bibitem{2005Chodorowski}
{Chodorowski} M~J 2005 {Cosmology Under Milne's Shadow} {\em \pasa\/} {\bf 22},
  287 (\textit{Preprint} \eprint{astro-ph/0503690})

\bibitem{Adamek2014}
{Adamek} J, {Di Dio} E, {Durrer} R and {Kunz} M 2014 {Distance-redshift
  relation in plane symmetric universes} {\em \prd\/} {\bf 89}(6) 063543
  (\textit{Preprint} \eprint{1401.3634})

\bibitem{Brynjolfsson2004}
{Brynjolfsson} A 2004 {Redshift of photons penetrating a hot plasma} {\em
  arXiv:astro-ph/0401420\/}

\bibitem{Brynjolfsson2006}
{Brynjolfsson} A 2006 {Magnitude-Redshift Relation for SNe Ia, Time Dilation,
  and Plasma Redshift} {\em ArXiv:astro-ph/0602500\/}

\bibitem{Marmet2018}
{Marmet} L 2018 {On the Interpretation of Spectral Red-Shift in Astrophysics: A
  Survey of Red-Shift Mechanisms - II} {\em arXiv e-prints\/} arXiv:1801.07582
  (\textit{Preprint} \eprint{1801.07582})

\bibitem{Zaninetti2015a}
{Zaninetti} L 2015 {On the Number of Galaxies at High Redshift} {\em
  Galaxies\/} {\bf 3}, 129

\bibitem{press}
{Press} W~H, {Teukolsky} S~A, {Vetterling} W~T and {Flannery} B~P 1992 {\em
  {Numerical Recipes in FORTRAN. The Art of Scientific Computing}\/}
  (Cambridge, UK: Cambridge University Press)

\bibitem{Akaike1974}
{Akaike} H 1974 {A new look at the statistical model identification} {\em IEEE
  Transactions on Automatic Control\/} {\bf 19}, 716

\bibitem{Liddle2004}
{Liddle} A~R 2004 {How many cosmological parameters?} {\em \mnras\/} {\bf 351},
  L49

\bibitem{Godlowski2005}
{Godlowski} W and {Szydowski} M 2005 {Constraints on Dark Energy Models from
  Supernovae} in M~{Turatto}, S~{Benetti}, L~{Zampieri} and W~{Shea}, eds, {\em
  1604-2004: Supernovae as Cosmological Lighthouses\/} (Astronomical Society of
  the Pacific) vol 342 of {\em Astronomical Society of the Pacific Conference
  Series\/} pp 508--516

\bibitem{bevington2003}
{{Bevington}, P~R and {Robinson}, D~K} 2003 {\em {Data Reduction and Error
  Analysis for the Physical Sciences}\/} (New York: McGraw-Hill)

\bibitem{Wei2010}
{Wei} H 2010 {Observational constraints on cosmological models with the updated
  long gamma-ray bursts} {\em \jcap\/} {\bf 8} 020 (\textit{Preprint}
  \eprint{1004.4951})

\bibitem{Etherington1933}
{Etherington} I~M~H 1933 {On the Definition of Distance in General Relativity.}
  {\em Philosophical Magazine\/} {\bf 15}

\bibitem{Remez1934}
{Remez} E 1934 Sur la d{\'e}termination des polyn{\^o}mes d{\'~}approximation
  de degr{\'e} donn{\'e}e {\em Comm. Soc. Math. Kharkov\/} {\bf 10}, 41

\bibitem{Remez1957}
Remez E 1957 {\em General Computation Methods of {C}hebyshev Approximation.
  {T}he Problems with Linear Real Parameters\/} (Kiev: Publishing House of the
  Academy of Science of the Ukrainian SSR)

\bibitem{Braatz2010}
{Braatz} J~A, {Reid} M~J, {Humphreys} E~M~L, {Henkel} C, {Condon} J~J and {Lo}
  K~Y 2010 {The Megamaser Cosmology Project. II. The Angular-diameter Distance
  to UGC 3789} {\em \apj\/} {\bf 718}(2), 657 (\textit{Preprint}
  \eprint{1005.1955})

\bibitem{Kuo2013}
{Kuo} C~Y, {Braatz} J~A, {Reid} M~J, {Lo} K~Y, {Condon} J~J, {Impellizzeri}
  C~M~V and {Henkel} C 2013 {The Megamaser Cosmology Project. V. An
  Angular-diameter Distance to NGC 6264 at 140 Mpc} {\em \apj\/} {\bf 767}(2)
  155 (\textit{Preprint} \eprint{1207.7273})

\bibitem{Melia2018}
{Melia} F and {Yennapureddy} M~K 2018 {The maximum angular-diameter distance in
  cosmology} {\em \mnras\/} {\bf 480}(2), 2144 (\textit{Preprint}
  \eprint{1807.07548})

\bibitem{Eales2010}
{Eales} S, {Dunne} L, {Clements} D and {Cooray} A 2010 {The Herschel ATLAS}
  {\em \pasp\/} {\bf 122}, 499 (\textit{Preprint} \eprint{0910.4279})

\bibitem{Tamura2015}
{Tamura} Y, {Oguri} M, {Iono} D, {Hatsukade} B, {Matsuda} Y and {Hayashi} M
  2015 {High-resolution ALMA observations of SDP.81. I. The innermost mass
  profile of the lensing elliptical galaxy probed by 30 milli-arcsecond images}
  {\em \pasj\/} {\bf 67} 72 (\textit{Preprint} \eprint{1503.07605})

\bibitem{ALMA2015}
{ALMA Partnership}, {Vlahakis} C, {Hunter} T~R and {Hodge} J~A 2015 {The 2014
  ALMA Long Baseline Campaign: Observations of the Strongly Lensed
  Submillimeter Galaxy HATLAS J090311.6+003906 at z = 3.042} {\em \apjl\/} {\bf
  808} L4 (\textit{Preprint} \eprint{1503.02652})

\bibitem{Rybak2015}
{Rybak} M, {Vegetti} S, {McKean} J~P, {Andreani} P and {White} S~D~M 2015 {ALMA
  imaging of SDP.81 - II. A pixelated reconstruction of the CO emission lines}
  {\em \mnras\/} {\bf 453}, L26 (\textit{Preprint} \eprint{1506.01425})

\bibitem{Hatsukade2015}
{Hatsukade} B, {Tamura} Y, {Iono} D, {Matsuda} Y, {Hayashi} M and {Oguri} M
  2015 {High-resolution ALMA observations of SDP.81. II. Molecular clump
  properties of a lensed submillimeter galaxy at z = 3.042} {\em \pasj\/} {\bf
  67} 93 (\textit{Preprint} \eprint{1503.07997})

\bibitem{Wong2015}
{Wong} K~C, {Suyu} S~H and {Matsushita} S 2015 {The Innermost Mass Distribution
  of the Gravitational Lens SDP.81 from ALMA Observations} {\em \apj\/} {\bf
  811} 115 (\textit{Preprint} \eprint{1503.05558})

\bibitem{Hezaveh2016}
{Hezaveh} Y~D, {Dalal} N and {Marrone} D~P 2016 {Detection of Lensing
  Substructure Using ALMA Observations of the Dusty Galaxy SDP.81} {\em \apj\/}
  {\bf 823} 37 (\textit{Preprint} \eprint{1601.01388})

\bibitem{Zaninetti2017c}
{Zaninetti} L 2017 The ring produced by an extra-galactic superbubble in flat
  cosmology {\em Journal of High Energy Physics, Gravitation and Cosmology\/}
  {\bf 3}, 339

\bibitem{CODATA2012}
{Mohr} P~J, {Taylor} B~N and {Newell} D~B 2012 {CODATA recommended values of
  the fundamental physical constants: 2010} {\em Reviews of Modern Physics\/}
  {\bf 84}, 1527

\bibitem{Pade1892}
{Pad{\'e}} H 1892 Sur la repr{\'e}sentation approch{\'e}e d'une fonction par
  des fractions rationnelles {\em Ann. Sci. Ecole Norm. Sup.\/} {\bf 9}, 193

\bibitem{Baker1975}
{Baker} G 1975 {\em Essentials of Pad{\'e} approximants\/} (New York: Academic
  Press)

\bibitem{Baker1996}
{Baker} G~A and {Graves-Morris} P~R 1996 {\em Pad{\'e} approximants\/} vol~59
  (Cambridge: Cambridge University Press)

\end{thebibliography}
\providecommand{\newblock}{}

\end{document}